\documentclass[lettersize,journal]{IEEEtran}
\usepackage{amsmath,amssymb,amsfonts}
\usepackage{algorithmic}
\usepackage{algorithm}
\usepackage{array}
\usepackage[caption=false,font=normalsize,labelfont=sf,textfont=sf]{subfig}
\usepackage{textcomp}
\usepackage{stfloats}
\usepackage{url}
\usepackage{verbatim}
\usepackage{graphicx}
\usepackage{cite}
\usepackage{xcolor}
\usepackage{bm}
\usepackage{bbm}
\usepackage{booktabs}
\usepackage[hidelinks]{hyperref}

\usepackage{orcidlink} 

\newcommand{\E}[1]{\mathcal{E}_{\text{#1}}}
\newcommand{\ind}[1]{\mathbbm{1}_{\{#1\}}}

\begin{document}

\title{Security Analysis of MDI-QKD in Turbulent Free-Space Polarization Channels — A Composite Channel Framework}
\author{Heyang~Peng\orcidlink{0000-0001-2345-6789},
        Seid~Koudia\orcidlink{0000-0002-3456-7890},
        and~Symeon~Chatzinotas\orcidlink{0000-0003-4567-8901},~\IEEEmembership{Fellow,~IEEE}}

\maketitle
\begin{abstract}
Atmospheric turbulence poses a significant challenge to free-space measurement-device-independent quantum key distribution (FSO MDI-QKD) by inducing polarization decoherence and depolarization, which degrade the secret key rate (SKR). In this paper, we propose a unified depolarizing–dephasing channel model for turbulence-induced polarization decoherence in FSO MDI-QKD. This model consolidates phase perturbations, Gaussian beam spreading, beam drift, aperture truncation, and scintillation into closed-form parameters: depolarization factor ($\lambda_a^{\text{eff}}$), decoherence factor ($r_a^{2,\text{eff}}$), and detection probability ($\eta_{\text{eff}}$). By mapping turbulence to a von Mises–Fisher/Watson-distributed SU(2) rotation, we derive an analytic SKR expression compatible with existing MDI-QKD security analyses. The model excels in clear, overcast, and hazy weather conditions, offering computational efficiency and experimental verifiability for real-time link adaptation. Numerical simulations, illustrated on a ground-to-satellite free-space link confirm its accuracy, enabling robust physical layer design for global-scale MDI-QKD networks

\end{abstract}
\section{Introduction}\label{sec:intro}
Quantum key distribution (QKD) relies on quantum measurement principles and the no-cloning theorem to provide information-theoretic security \cite{b1,b2}. Measurement-device-independent QKD (MDI-QKD) strengthens this by employing an untrusted relay for the Bell-state measurements (BSM), mitigating all detector side-channel attacks \cite{b3,b4}. To enable global-scale quantum communication links, free-space optical (FSO) links are attractive due to their low geometric loss and with reduced terrestrial infrastructure compared to long-haul fiber-based intercity QKD, only endpoint optical ground stations are required on the ground\cite{b5,b6,b7,b8,b9,b10,b11,b12}. Atmospheric turbulence, however, introduces scintillation, beam drift, and wavefront distortions that can induce polarization mixing and degrade the secret key rate (SKR) \cite{b13,b14,b15,b16,b17,b18,b19}. Classical statistical optics connects these impairments to path-integrated index fluctuations and aperture averaging, while quantum-channel descriptions clarify how turbulence modifies photonic qubits and link observables \cite{b15,b16,b17}. Field measurements further indicate that polarization can remain largely preserved on space-to-ground paths, though measurable depolarization may arise depending on geometry and weather \cite{b18}.

Previous work on turbulence effects falls short for MDI-QKD applications. Simplified models often reduce the turbulence to a single misalignment angle, which leads to optimistic assumptions about the stability of the polarization \cite{b20, b21, b22, b23}. For example, adaptive optics (AO) and single-mode filtering stabilize short-range links but do not capture complex polarization dynamics under diverse weather conditions \cite{b8,b16,b24}. Conversely, Monte Carlo wave optics simulations, including split-step propagation methods, provide detailed wavefront analysis, yet incur high computational costs, scaling with propagation distance and aperture size, making them computationally prohibitive for real-time optimization \cite{b25, b26, b27, b28, b29}. Recent advances, such as mode diversity, quantum MIMO, and continuous-variable or entanglement-based protocols, improve resilience in specific settings but still lack a compact, polarization-qubit–oriented description that integrates directly into MDI-QKD security analyzes \cite{b30,b31,b32,b33,b34,b35,b36,b37,b38}. These approaches do not offer closed-form parameters directly compatible with MDI-QKD security proofs, such as those for thermal-loss and phase-noise channels \cite{b38}, nor do they fully account for beam propagation through random media or quantum repeaters in space-based networks \cite{b30,b36,b37,b38}.

We propose a composite quantum channel model that unifies turbulence-induced effects: phase perturbations, diffraction-induced Gaussian beam spreading, beam drift, aperture truncation, and scintillation in a depolarizing-dephasing framework. By modeling polarization rotations with SU(2) unitaries whose axes follow a von Mises–Fisher/Watson distribution, we derive closed-form expressions for the depolarization factor ($\lambda_a^{\text{eff}}$), decoherence factor ($r_a^{2,\text{eff}}$), and detection probability ($\eta_{\text{eff}}$), seamlessly integrating with security analyzes \cite{b38}. This model enables analytic SKR predictions, is computationally lightweight, and is amenable to real-time link adaptation for terrestrial and satellite links under clear, overcast, and hazy conditions.

We focus on the weak-turbulence regime (\(\sigma_R^2 \le 1\)); unless otherwise stated, all derivations and quantitative claims are calibrated in this regime. 
For completeness, we also report a phenomenological extension to medium turbulence (\(1<\sigma_R^2<5\)) using a saturated phase-structure function, used only for trend illustration, as well as strong turbulence (\(\sigma_R^2>5\)). The three weather labels, clear, overcast, and hazy, denote different Hufnagel-Valley (HV) parameter sets within this regime. We also assume a polarization-insensitive receiver and near-axis propagation. These choices are consistent with the targeted ground-to-satellite scenarios and with our validation range. Our numerical study focuses on uplinks; the analytical channel is geometry-agnostic and applies to both directions. 

This paper is organized as follows. Section~\ref{sec:channel} formulates the composite channel model; Section~\ref{sec:turbulence_regimes} derives regime-specific parameters; Section~\ref{sec:security} integrates these into MDI-QKD security analysis for SKR; Section~\ref{sec:simulations} validates the model through numerical simulations; and Section~\ref{sec:conclusion} discusses implications and future work.

\section{Quantum Channel Framework for Polarization Evolution}\label{sec:channel}

\begin{figure}[!t]
\centering
\includegraphics[width=\columnwidth]{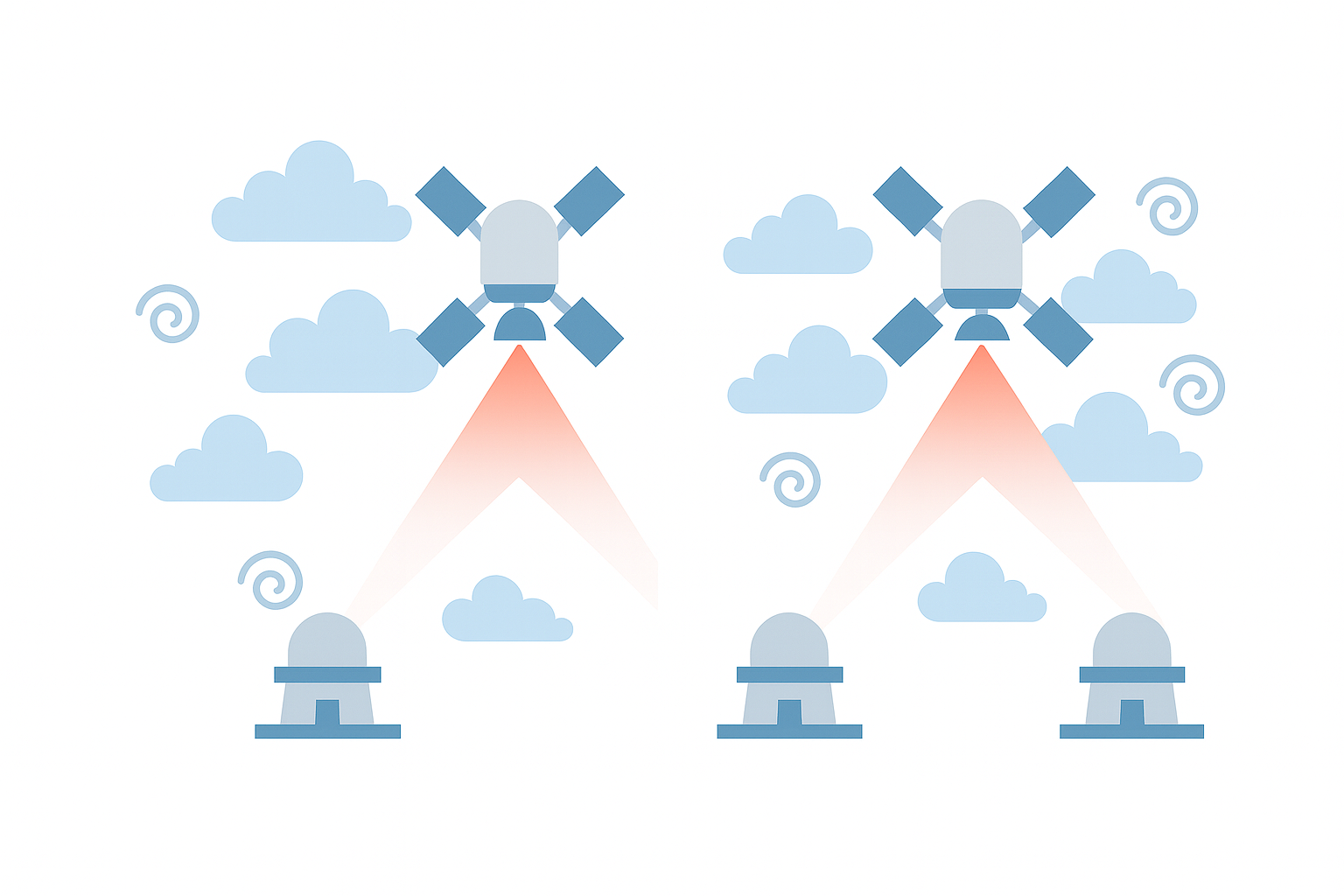}  
\caption{Schematic of the ground-to-satellite FSO MDI-QKD link. }  
\label{fig:system_schematic}
\end{figure}

In FSO MDI-QKD (schematic in Fig.~\ref{fig:system_schematic}), polarization states serve as the primary carriers of quantum information. However, atmospheric turbulence introduces complex perturbations that degrade these states, challenging the fidelity and security of the protocol. To understand these effects, we first model the photon's wavefront as a Gaussian beam, which approximates the transverse mode of the laser source and provides an analytical basis for propagation analysis. The electric field of such a beam, propagating along the \(z\)-axis, is given by
\begin{equation}
\label{eq:gaussian_field}
\begin{aligned}
E(\vec{\rho}, z)
&= E_0\,\frac{w_0}{w_z}\,
\exp\!\Big(-\frac{|\vec{\rho}|^2}{w_z^2}\Big) \\
&\quad\times
\exp\!\Big(i k z + i \frac{k |\vec{\rho}|^2}{2 R_z} - i \zeta(z) + i \phi_{\text{turb}}(\vec{\rho}, z)\Big).
\end{aligned}
\end{equation} 
where \(E_0\) is the initial field amplitude; \(w_z = w_0 \sqrt{1 + (z/z_R)^2}\) is the radius of the beam in the range \(z\) with \(w_0\) the waist of the beam and \(z_R = \pi w_0^2 / \lambda\) the Rayleigh range; \(k = 2\pi / \lambda\) is the wavenumber; \(R_z = z \big( 1 + (z_R/z)^2 \big)\) is the radius of curvature; \(\zeta(z) = \arctan(z/z_R)\) is the Gouy phase; and \(\phi_{\text{turb}}(\vec{\rho}, z)\) is the phase induced by turbulence. We write \(n(\vec{\rho},z)=n_0+\delta n(\vec{\rho},z)\) with zero-mean fluctuations \(\mathbb{E}[\delta n]=0\), so \(\mathbb{E}[\phi_{\text{turb}}]=0\).

Without turbulence, the Gaussian beam evolves predictably through diffraction and phase accumulation, maintaining coherence. However, atmospheric turbulence disrupts this by introducing random phase fluctuations \(\phi_{\text{turb}}(\vec{\rho}, z)\), resulting from spatial and temporal variations in the refractive index \(n(\vec{\rho}, z)\), primarily caused by temperature gradients, pressure changes, and wind mixing. These aberrations accumulate along the path as
\begin{equation}
\phi_{\text{turb}}(\vec{\rho}, z) = k \int_0^z \delta n(\vec{\rho}, \xi) d\xi,
\end{equation}
Here, \(\delta n = n - n_0\) denotes fluctuations in the zero mean index after removing the mean refractive index \(n_0\),  where \(\delta n\) follows statistical models like the von Kármán spectrum (see Appendix~\ref{app:turbulence_models}). Small-scale eddies (inner scale \(l_0 \sim 1~\text{mm}\), scenario-dependent) ,  large-scale eddies (outer scale \(L_0 \sim 25~\text{m}\)) cause beam drift and tilt.

For polarization qubits, these phase gradients can act as an effective birefringence once combined with the optical train (e.g., PBS/fiber coupling), slanted geometry, and aperture-weighted spatial averaging, resulting in random SU(2) rotations on the polarization Bloch sphere that cause depolarization (mixing toward \(\mathbb{I}/2\)) and decoherence (damping off-diagonal terms).
To systematically address these challenges, we develop a comprehensive quantum channel model that captures the impact of turbulence on the polarization degree of freedom. Our approach consolidates the effects—phase perturbations, Gaussian beam spreading driven by diffraction, beam drift, aperture truncation, and scintillation—into a physically motivated sequence of transformations; the turbulence spectra and path-averaged strength follow the von Kármán and HV models summarized in Appendix~\ref{app:turbulence_models} and Table~\ref{tab:params}.

\subsection{Overview of the Composite Channel}
The composite quantum channel that acts on the polarization density matrix $\hat{\rho}$ is defined as
\begin{equation}
\mathcal{E}_{\text{composite}} = \mathcal{E}_{\text{aperture}} \circ \mathcal{E}_{\text{drift}} \circ \mathcal{E}_{\text{spread}} \circ \mathcal{E}_{\text{phase}}.
\end{equation}
We follow the standard phase-screen (split-step) picture and thus organize the propagation as follows:
phase perturbations ($\mathcal{E}_{\text{phase}}$) model random polarization rotations from refractive index variations; 
Gaussian beam spreading ($\mathcal{E}_{\text{spread}}$) accounts for diffraction-induced spatial distribution; 
beam drift ($\mathcal{E}_{\text{drift}}$) captures large-scale displacements; 
and aperture truncation ($\mathcal{E}_{\text{aperture}}$) limits collection to the receiver's finite area (applied at the receiver plane). 
This ordering is a modeling device rather than a physical time order; in the infinitesimal-step limit, it is equivalent to simultaneous action of turbulence. Scintillation, modeled separately, affects the probability of detection $\eta_{\text{eff}}$, but not the polarization state. 
The detected (post-selected) polarization state and efficiency are then given by

\begin{align}
\hat{\rho}' \; &= \;
\frac{{\rm Tr}_{\rm spatial}\!\big[\Pi_a\,\mathcal{E}_{\text{drift}}\!\circ\!\mathcal{E}_{\text{spread}}\!\circ\!\mathcal{E}_{\text{phase}}(\hat{\rho})\,\Pi_a\big]}
{{\rm Tr}\!\big[\Pi_a\,\mathcal{E}_{\text{drift}}\!\circ\!\mathcal{E}_{\text{spread}}\!\circ\!\mathcal{E}_{\text{phase}}(\hat{\rho})\,\Pi_a\big]},
\\
\eta_{\text{eff}} \; &= \;
{\rm Tr}\!\big[\Pi_a\,\mathcal{E}_{\text{drift}}\!\circ\!\mathcal{E}_{\text{spread}}\!\circ\!\mathcal{E}_{\text{phase}}(\hat{\rho})\,\Pi_a\big].
\end{align}
${\rm Tr}_{\rm spatial}[\cdot]$ denotes the partial trace over the spatial modes of the receiver plane and ${\rm Tr}[\cdot]$ the full trace. Thus, the overall map on the joint system of spatial degrees of freedom and polarization is completely positive and trace preserving (CPTP); conditioning on detection induces a completely positive trace non increasing (CPTNI) map on polarization that is normalized to CPTP.

The strength of turbulence is quantified by the Rytov variance \(\sigma_R^2\) (Appendix~\ref{app:turbulence_models}), categorizing regimes: weak (\(\sigma_R^2 < 1\)), medium (\(1 < \sigma_R^2 < 5\)), and strong (\(\sigma_R^2 > 5\)). Building on this general framework, we detail each component in the following.
The composite map assumes (i) scalar, polarization-insensitive intensity fluctuations and a nonpolarizing receiver (no diattenuation), (ii) paraxial Gaussian beams with path-averaged $C_n^2$ and von Karman spectra, and (iii) a detection-conditioned (post-selected) polarization map that is CPTNI but normalized to CPTP for state analysis. Potential weak diattenuation (e.g. due to aerosols or coatings) is neglected at first order in our weak-turbulence regime; a perturbative bound is deferred to the Appendix~\ref{app:scintillation}. Weather labels (clear/overcast/hazy) instantiate distinct weak-regime HV profiles without adding aerosol extinction; when present, aerosols primarily rescale $\eta_{\rm total}$ via $\alpha_{\rm atm}$ with negligible impact on $(\lambda_a^{\rm eff},r_a^{2,{\rm eff}})$.
\subsection{Phase Channel: From Wavefront Phase to Polarization Rotations}
\label{sec:phase_channel}
\begin{figure}[!t]
\centering
\includegraphics[width=\columnwidth]{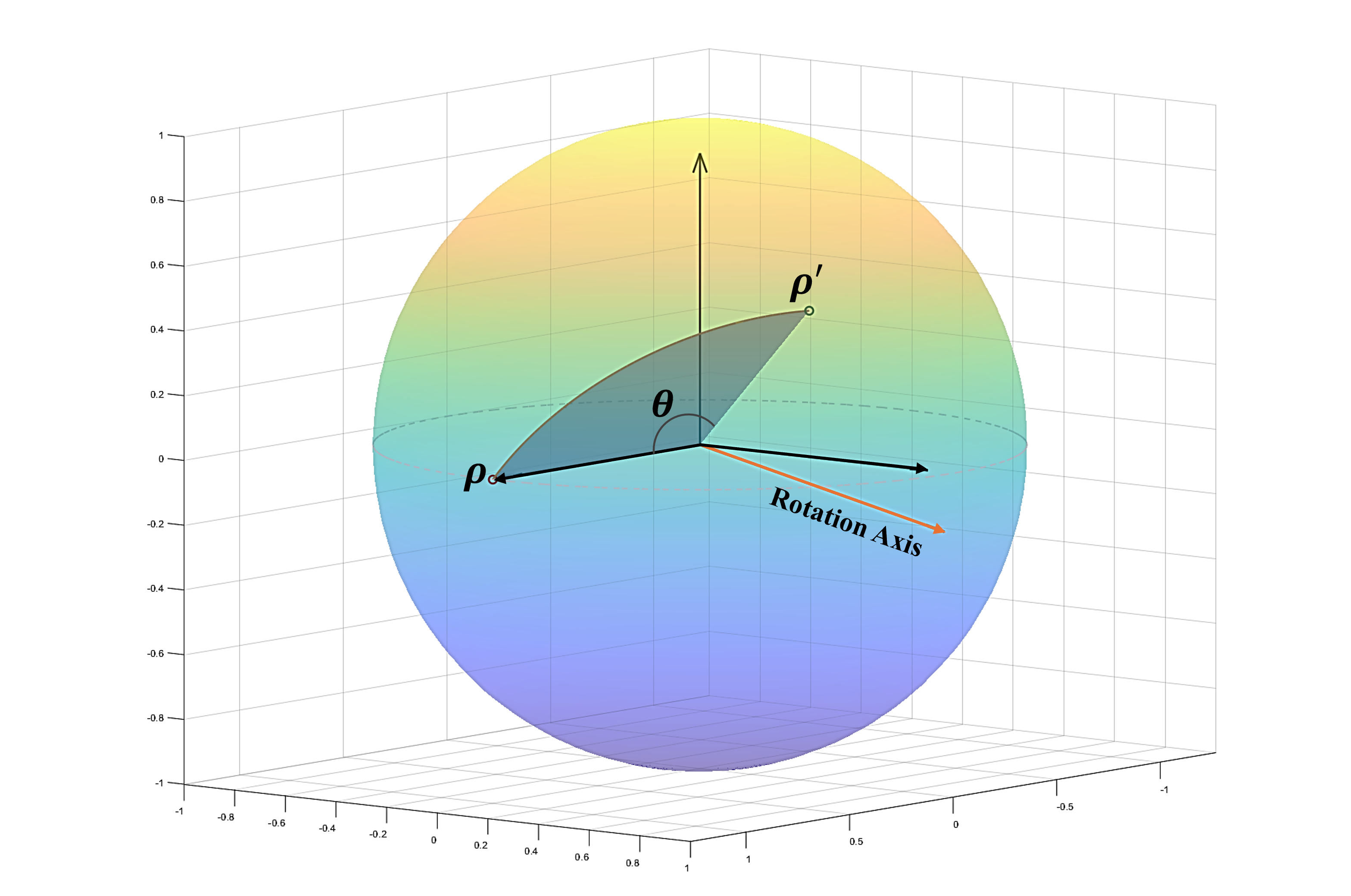}
\caption{Illustration of turbulence-induced polarization rotation on the Bloch sphere. An initial polarization state $\rho$ is rotated by angle $\theta$ about a random axis, resulting in the state $\rho'$. }
\label{fig:bloch_sphere}
\end{figure}

Building on the turbulence-induced phase term \(\phi_{\text{turb}}(\vec{\rho}, z)\) in the Gaussian beam field, these random wavefront distortions translate into stochastic rotations of the polarization state within the Stokes space. We model this using the density matrix formalism, which effectively captures depolarization, mixing toward the identity matrix \(\mathbb{I}/2\) and decoherence, suppression of off-diagonal elements, arising from phase averaging.
Fig.~\ref{fig:bloch_sphere} shows the Bloch sphere representation of polarization rotations.

The polarization rotation is represented by a unitary operator in the SU(2) group:
\begin{equation}
U(\theta, \vec{n}) = \cos\!\left(\frac{\theta}{2}\right)\mathbb{I}
- i \sin\!\left(\frac{\theta}{2}\right) (\vec{n} \cdot \vec{\sigma}),
\end{equation}
where \(\vec{n}=(\sin\vartheta\cos\varphi,\;\sin\vartheta\sin\varphi,\;\cos\vartheta)\) denotes the rotation axis in the Stokes space, \(\vec{\sigma} = (\sigma_x, \sigma_y, \sigma_z)\) are the Pauli matrices, \(\theta\) is the rotation angle and \(\vartheta\), \(\varphi\) are the polar and azimuthal angles, respectively.
To encode slant-path anisotropy with axial symmetry about the line of sight, we model the SU(2) rotation axis as an axial Watson/von–Mises–Fisher distribution around the propagation direction\cite{b29}:
\begin{equation}
w_\kappa(\vartheta)=\frac{\kappa}{2\sinh\kappa}e^{\kappa\cos\vartheta}\sin\vartheta,\ \kappa\ge 0.
\end{equation}
Here, \(\kappa\) serves as the concentration parameter, confining axes around the \(+z\) direction; this distribution is particularly suitable as it models directional preferences in turbulent media, transitioning from isotropic (\(\kappa=0\)) to aligned (\(\kappa \to \infty\)) behaviors. 
The rotation angle \(\theta\) obeys a conditional distribution \(p(\theta \mid r)\), with \(r = |\vec{\rho}|\) as the radial distance from the beam center; this distribution adapts to the turbulence regime (detailed in Section~\ref{sec:turbulence_regimes} and Appendix~\ref{app:phase_derivation}).
The Kraus operators for the phase channel are given by
\begin{equation}
\begin{aligned}
    K_{\theta,\vartheta,\varphi} \;&=\;
\sqrt{ p_\theta(\theta\mid r)\, w_\kappa(\vartheta)\,\frac{\sin\vartheta}{2\pi} } \;\, U(\theta,\vec n),\\
\qquad
U(\theta,\vec n)&=\exp\!\big[-\tfrac{i}{2}\theta\,(\vec n\!\cdot\!\vec\sigma)\big],
\end{aligned}
\end{equation}
where $\theta\!\in[0,\pi]$, $(\vartheta,\varphi)\!\in[0,\pi]\!\times\![0,2\pi)$ parameterize the rotation axis $\vec n$ on the sphere, $p_\theta(\cdot\mid r)$ and $w_\kappa(\cdot)$ are normalized densities.
\begin{equation}
\mathcal{E}_{\text{phase},\,\vec{\rho}}(\hat{\rho})
=\!\!\int_{0}^{\pi}\!\!\!\mathrm d\theta
\int_{0}^{\pi}\!\!\!\mathrm d\vartheta
\int_{0}^{2\pi}\!\!\!\mathrm d\varphi\;
K_{\theta,\vartheta,\varphi}\;\hat{\rho}\;K_{\theta,\vartheta,\varphi}^\dagger .
\label{eq:phase_channel_kraus}
\end{equation}
This operation leads to a depolarizing--dephasing channel, mixing the state toward $\mathbb{I}/2$ while damping coherences. The effects are quantified by the depolarization factor $\lambda(r)$ and the decoherence factor $\bar{r}^2(r, \kappa)$, obtained by averaging the rotation parameters (see Appendix~\ref{app:phase_derivation} for the derivations).  Specifically, the channel takes the form:
\begin{equation}
\mathcal{E}_{\text{phase},\,\vec{\rho}}(\hat{\rho}) =
(1 - \lambda(r))
\begin{bmatrix}
\rho_{HH} & \rho_{HV} \bar{r}^2(r, \kappa) \\
\rho_{VH} \bar{r}^2(r, \kappa) & \rho_{VV}
\end{bmatrix}
+ \frac{\lambda(r)}{2} \mathbb{I}.
\label{eq:phase_channel}
\end{equation}
These factors are measurable via polarimetric experiments or data fitting, supporting practical implementation and maintaining quantum mechanical consistency.
\subsection{Gaussian Beam Spreading}
\label{sec:gaussian_beam}

Following the position-dependent phase perturbations modeled in the previous subsection, the Gaussian beam spreading channel \(\mathcal{E}_{\text{spread}}\) incorporates the spatial extent of the beam, averaging these localized effects across the receiver plane. Gaussian beams are chosen for their well-defined intensity profile and analytical tractability, approximating the transverse mode of the photon's wavefront with a Gaussian transverse distribution.

The intensity profile at the receiver plane is given by
\begin{equation}
I(\vec{\rho}) = \frac{2}{\pi w_z^2}\exp\!\left(-\frac{2\lvert\vec{\rho}\rvert^2}{w_z^2}\right),
\label{eq:I_zero_drift}
\end{equation}
where \(w_z\) denotes the radius of the beam after propagation, influenced by diffraction over distances much larger than the Rayleigh range (\(z \gg z_R\)). This spreading distributes the photon's probability density, weighting turbulence-induced phase rotations by the local intensity.

The spreading channel is defined with the Kraus operators:
\begin{equation}
K_{\vec{\rho}} = \sqrt{I(\vec{\rho})}\, |\vec{\rho}\rangle \langle \vec{\rho}| \otimes \mathbb{I}_{\text{pol}},
\end{equation}
where \(\mathbb{I}_{\text{pol}}\) preserves the polarization subspace. The channel acts on a state as
\begin{equation}
\mathcal{E}_{\text{spread}}(\hat{\rho}) = \int d^2\vec{\rho}\, K_{\vec{\rho}}\, \mathcal{E}_{\text{phase},\,\vec{\rho}}(\hat{\rho})\, K_{\vec{\rho}}^\dagger.
\end{equation}
By integrating over the Gaussian profile, this operation spatially averages the phase-perturbed state, with central regions (higher intensity) dominating the contribution. This coupling is essential, as it links varying phase effects across the beam to the effective polarization at detection, ensuring that the model reflects realistic propagation dynamics.

\subsection{Beam Drift}
\label{sec:beam_drift}

Following the spatial averaging of phase perturbations over the Gaussian intensity profile, the beam drift channel \(\mathcal{E}_{\text{drift}}\) introduces random displacements of the entire beam due to large-scale turbulent eddies, further modulating the effective polarization state at the receiver.

Conditioned on a fixed lateral drift \(\vec r_{\text{drift}}\), the normalized spatial weight at the receiver is
\begin{equation}
\omega(\vec{\rho},\vec r_{\text{drift}})
= \frac{2}{\pi w_z^{2}} \exp\!\left(-\frac{2\,\lvert\vec{\rho}-\vec r_{\text{drift}}\rvert^{2}}{w_z^{2}}\right).
\label{eq:omega}
\end{equation}

Beam drift is modeled as a transverse shift \(\vec{r}_{\text{drift}} = (x_{\text{drift}}, y_{\text{drift}})\) with a Gaussian probability distribution:
\begin{equation}
p(\vec{r}_{\text{drift}}) = \frac{1}{2\pi \sigma_{\text{drift}}^2} \exp\left(-\frac{|\vec{r}_{\text{drift}}|^2}{2 \sigma_{\text{drift}}^2}\right),
\end{equation}
where the variance \(\sigma_{\text{drift}}^2\) is derived from the von Kármán tilt spectrum under the spherical-wave approximation (see Appendix~\ref{app:turbulence_models} for derivation):
\begin{equation}
\sigma_{\text{drift}}^2 = 0.73\, \langle C_n^2 \rangle\, z^3\, w_z^{-1/3} \left[1 - 0.97 \left(\frac{L_0}{w_z}\right)^{1/3}\right] \sec^2\beta.
\end{equation}
Here, \(L_0 = 25\,\text{m}\) is the outer scale of turbulence, \(\langle C_n^2 \rangle\) is the path-averaged refractive index structure constant, and \(\beta\) is the zenith angle. The prefactor 0.73 reflects spherical-wave geometry, the correction term ensures positive variance by accounting for averaging over large eddies when \(w_z > L_0\), and \(\sec^2\beta\) adjusts for extended path lengths in slanted propagation.

The Kraus operators for the drift channel are
\begin{equation}
K_{\vec{r}_{\text{drift}}} = \sqrt{p(\vec{r}_{\text{drift}})}\, |\vec{r}_{\text{drift}}\rangle \langle \vec{r}_{\text{drift}}| \otimes \mathbb{I}_{\text{pol}},
\end{equation}
defining the channel as
\begin{equation}
\mathcal{E}_{\text{drift}}(\hat{\rho}) = \int d^2 \vec{r}_{\text{drift}}\, K_{\vec{r}_{\text{drift}}}\, \mathcal{E}_{\text{spread},\,\vec{\rho} - \vec{r}_{\text{drift}}}(\hat{\rho})\, K_{\vec{r}_{\text{drift}}}^\dagger.
\end{equation}

This operation shifts the spatial coordinates of the spread state by \(\vec{r}_{\text{drift}}\), preserving polarization but repositioning the beam relative to the aperture. The shift couples with the Gaussian intensity, influencing detection probability and polarization averaging. In severe cases, large drifts may cause signal loss if the aperture limits are exceeded, underscoring the role of adaptive optics in mitigating such effects.

\subsection{Scintillation}
\label{sec:scintillation}
Complementing the spatial filtering from prior channels, scintillation accounts for turbulence-induced focusing/defocusing due to distorted wavefront interference. We model the intensity multiplier $I_0$ with the Gamma–Gamma law
\begin{equation}
\resizebox{\columnwidth}{!}{$
p(I_0)=
\frac{2\,(\alpha_{\rm GG}\beta_{\rm GG})^{\frac{\alpha_{\rm GG}+\beta_{\rm GG}}{2}}}
{\Gamma(\alpha_{\rm GG})\Gamma(\beta_{\rm GG})}\,
I_0^{\frac{\alpha_{\rm GG}+\beta_{\rm GG}}{2}-1}\,
K_{\alpha_{\rm GG}-\beta_{\rm GG}}\!\left(2\sqrt{\alpha_{\rm GG}\beta_{\rm GG} I_0}\right)
$}
\label{eq:gamma_gamma_body}
\end{equation}

with shape parameters (standard aperture-averaged fits)
\begin{align}
\alpha_{\rm GG}&=\!\left[\exp\!\left(\frac{0.49\,\sigma_R^2}{\big(1+0.18 d^2+0.56\sigma_R^{12/5}\big)^{7/6}}\right)\!-1\right]^{-1},\nonumber\\
\beta_{\rm GG}&=\!\left[\exp\!\left(\frac{0.51\,\sigma_R^2(1+0.69\sigma_R^{12/5})^{-5/6}}{1+0.9 d^2+0.62 d^2\sigma_R^{12/5}}\right)\!-1\right]^{-1},\nonumber\\
d&=\sqrt{\frac{k a^2}{z}},\quad k=\tfrac{2\pi}{\lambda}.
\label{eq:alpha_beta_GG}
\end{align}
Under a polarization-insensitive receiver, linear detector response within the gate, and no polarization-dependent loss in post-turbulence optics, scintillation is a scalar modulation and leaves polarization unchanged:
\begin{equation}
\mathcal{E}_{\rm scint}(\hat{\rho})=\hat{\rho}.
\label{eq:scint_channel}
\end{equation}
Its role is confined to the detection probability,
\begin{equation}
\eta_{\rm eff}
=\int_{0}^{\infty}\!p(I_0)\,I_0\,\mathrm{d}I_0
\;\times\!
\int_{0}^{\infty}\!p_{\rm drift}(r_{\rm drift})\,\eta(r_{\rm drift})\,\mathrm{d}r_{\rm drift},
\label{eq:eta_eff_body}
\end{equation}
where $p_{\rm drift}(r_{\rm drift})$ is the radial pdf of beam drift and $\eta(r_{\rm drift})$ is the capture/throughput for a given drift (defined in the main text via the radial weight $W$). The cancellation of $I_0$ in the normalized density matrix, and thus the independence of $(\lambda_a^{\rm eff},\,r_a^{2,{\rm eff}})$ from scintillation, is shown in Appendix~\ref{app:scintillation}. In very strong turbulence, heavier-tailed alternatives (e.g., K-distribution) may improve tail fits, but they primarily affect $\eta_{\rm eff}$ rather than the polarization parameters.
\begin{figure}[!t]
\centering
\includegraphics[width=\columnwidth]{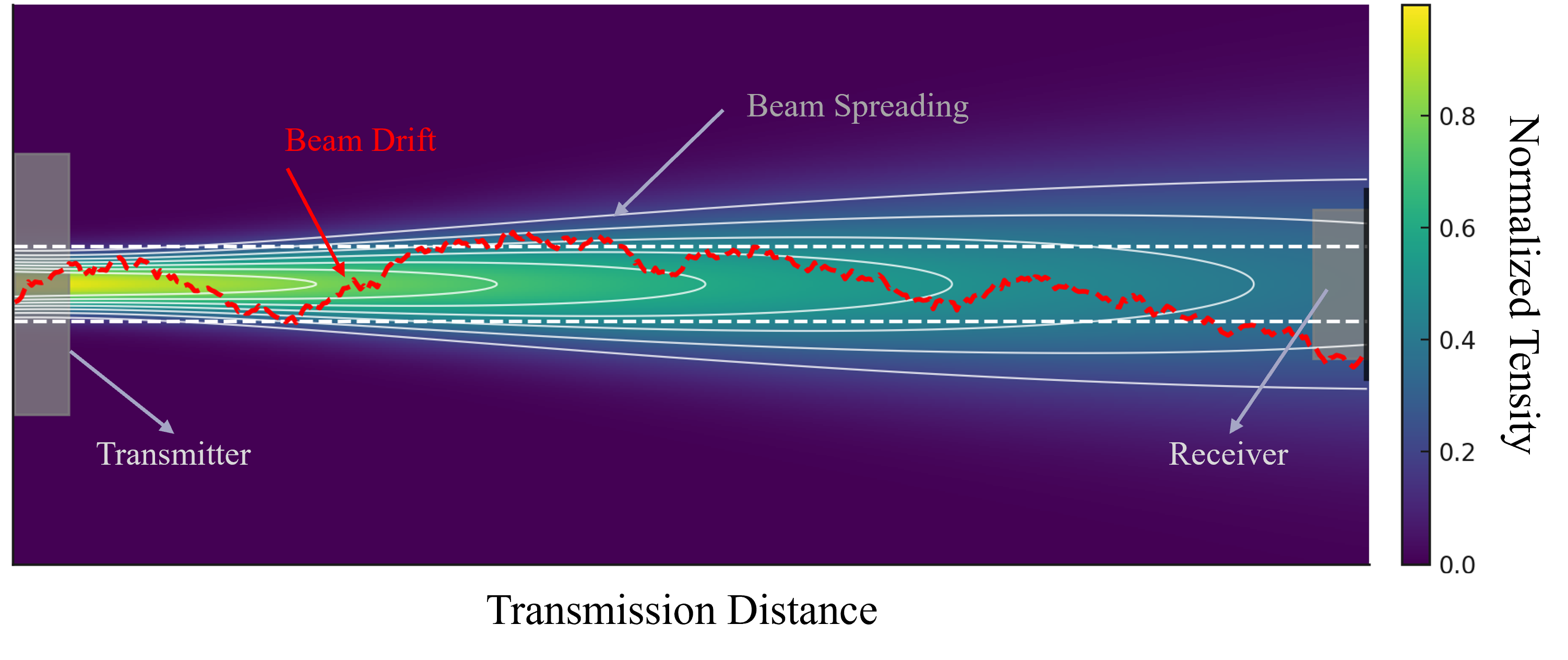}  
\caption{Illustration of beam spreading and beam drift in turbulent free-space propagation.}  
\label{fig:Beam_propagation}
\end{figure}

\subsection{Aperture Truncation}
\label{sec:aperture}

Building on the intensity fluctuations from the scintillation, the aperture truncation channel \(\mathcal{E}_{\text{aperture}}\) models the finite receiver collection, integrating these effects into the detection probability while finalizing the spatial filtering of the drifted beam.

The receiver’s finite aperture, with radius \(a = 0.5\,\text{m}\), collects only photons within \(|\vec{\rho}| \leq a\). The projection operator is
\begin{equation}
\Pi_a = \int_{|\vec{\rho}| \leq a} |\vec{\rho}\rangle \langle \vec{\rho}| \, d^2 \vec{\rho}.
\end{equation}
The Kraus operators are
\begin{align}
K_a &= \Pi_a \otimes \mathbb{I}_{\text{pol}}, \\
K_{a^\perp} &= (\mathbb{I}_{\text{spatial}} - \Pi_a) \otimes \mathbb{I}_{\text{pol}},
\end{align}
defining the channel as
\begin{equation}
\mathcal{E}_{\text{aperture}}(\hat{\rho}) = \text{Tr}_{\text{spatial}} \left[ \Pi_a\, \mathcal{E}_{\text{drift}}(\hat{\rho})\, \Pi_a \right] / \eta_{\text{eff}},
\end{equation}
where the detection probability incorporates scintillation via \(p(I_0)\) (detailed in Section~\ref{sec:scintillation} and Appendix~\ref{app:scintillation}):
\begin{equation}
\resizebox{\columnwidth}{!}{$
\eta_{\text{eff}}
= \int_{0}^{\infty} p(I_0)\, I_0
\bigg[ \int_{0}^{\infty} p_{\text{drift}}(r_{\text{drift}})
      \underbrace{\left(\int_{0}^{a} W(r,\,r_{\text{drift}})\,\mathrm{d}r\right)}_{\text{aperture capture}}
      \mathrm{d}r_{\text{drift}} \bigg] \mathrm{d}I_0
$}
\label{eq:eta_eff}
\end{equation}

The factor $(1-e^{-2a^2/w_z^2})$ is a small drift approximation to the Gaussian capture fraction. For general drifts, we evaluate the capture as
$\int_{0}^{a} W\!\left(r,\,r_{\text{drift}}\right)\,\mathrm{d}r$
using the normalized spatial weight \(\omega(\vec{\rho},\vec r_{\text{drift}})\) (Eq.~\eqref{eq:omega}) via its radial form 
\(W(r,\,r_{\text{drift}})=\int_{0}^{2\pi}\omega(\vec{\rho},\vec r_{\text{drift}})\,r\,\mathrm{d}\varphi\) (Eq.~\eqref{eq:Wradial}).
This leads to a numerically robust, CPTP-consistent averaging over the joint spatial and polarization degrees of freedom.
Here, \(r_{\text{drift}} = |\vec{r}_{\text{drift}}|\) is the radial displacement. The term \(1 - \exp(-2 a^2 / w_z^2)\) approximates the Gaussian fraction captured for small drifts, with numerical handling for larger ones.

This channel traces out uncollected modes, normalizing the polarization state. By selectively sampling perturbed regions from phase, spreading, and drift - modulated by scintillation intensity - it reshapes polarization (e.g., off-center drifts bias toward edges with unique phases) and sets transmission probability through geometric and fluctuation losses.

\subsection{Atmospheric Attenuation}
\label{sec:attenuation}

Complementing the intensity modulation from scintillation, atmospheric attenuation due to absorption and scattering (e.g., by molecules, aerosols, or water vapor) imposes a polarization-independent loss, modeled via the Beer--Lambert law as a scalar factor:
\begin{equation}
L_{\text{atm}}(\theta_{\rm el}) = \exp\left[ -\alpha_{\text{atm}} \, z_{\text{atm}}(\theta_{\rm el}) \right],
\end{equation}
where $z_{\text{atm}}(\theta_{\rm el}) = H_{\text{atm}} / \sin \theta_{\rm el}$, $\theta_{\rm el}$ is the elevation angle, and $H_{\text{atm}} = 20\,\text{km}$ is the effective atmospheric thickness, aligned with turbulence truncation at $20\,\text{km}$. The attenuation coefficient $\alpha$ varies with wavelength and conditions ---- for example, lower in the near-infrared (around $0.004\,\text{km}^{-1}$ on clear nights) and higher in visible bands or foggy weather (up to $0.02\,\text{km}^{-1}$ or more) ---- reflecting differences in molecular absorption, Rayleigh scattering, and aerosol extinction across spectra and turbulence levels.

For precise modeling, a height-dependent $\alpha_\text{atm
}(h)$ may be employed:
\begin{equation}
L_{\text{atm}}(\theta_{\rm el}) = \exp\left( -\int_0^z \alpha_{\text{atm}}(h(\xi))\, d\xi\, /\, \sin\theta_{\rm el} \right).
\end{equation}

A constant $\alpha_\text{atm}$ often suffices, as evidenced in satellite QKD studies~\cite{b6,b18}.

The total detection probability integrates this attenuation:
\begin{align}
\eta_{\text{total}} &= L_{\text{atm}}(\theta_{\rm el})\, \cdot\, \eta_{\text{eff}} \notag \\
&= L_{\text{atm}}(\theta_{\rm el})\! \int_0^\infty\! p(I_0)\, I_0 \biggl[ \int_0^\infty \frac{r_{\text{drift}}}{\sigma_{\text{drift}}^2} \exp\!\left( -\frac{r_{\text{drift}}^2}{2 \sigma_{\text{drift}}^2} \right) \notag \\
&\qquad \times \left( 1 - \exp\left( -\frac{2 a^2}{w_z^2} \right) \right) \ind{r_{\text{drift}} < a}\, dr_{\text{drift}} \biggr] dI_0. \label{eq:eta_total}
\end{align}
The total output state is:

\begin{equation} 
\hat{\rho}'_{\rm total}=\eta_{\rm total}\,\hat{\rho}'+(1-\eta_{\rm total})\,|\text{e}\rangle\!\langle\text{e}|.
\end{equation}

Here, $|\text{e}\rangle\!\langle\text{e}|$ is the vacumm state. This model accounts for environmental losses that are multiplicative and wavelength-dependent, simplifying analysis while preserving accuracy across varying atmospheric and spectral conditions.

\subsection{Turbulence Regimes and Polarization State}
\label{sec:turbulence_regimes}
Building on the general framework, we now derive regime-specific expressions for \(\lambda_a^{\text{eff}}\) and \(r_a^{2,\text{eff}}\). Having established the sequence of the composite channel and incorporated attenuation effects on the probability of detection, we examine how the phase perturbation channel \(\mathcal{E}_{\text{phase},\,\vec{\rho}}\) adapts to varying turbulence strengths, quantified by the Rytov variance \(\sigma_R^2\) (detailed in the Appendix~\ref{app:turbulence_models}). This variance classifies turbulence into regimes that dictate the statistical distribution of phase fluctuations, thereby influencing depolarization and decoherence.
von Mises–Fisher/Watson axis model introduces a single concentration parameter $\kappa$ to capture anisotropy along the propagation direction. Our use of $\kappa$ is phenomenological, yet CPTP-consistent via the Bloch shrinkage mapping; a practical calibration $\kappa(\sigma_R^2)$ may be obtained by fitting least squares to wave optic data sets or on-sky polarimetry. Likewise, the drift variance uses a spherical wave scaling with path-averaged $\langle C_n^2\rangle$; altitude-resolved $C_n^2(h)$ may refine $\sigma_{\rm drift}^2$ without altering the channel structure. These refinements do not change the closed-form interface $(\lambda_a^{\rm eff}, r_a^{2,{\rm eff}}, \eta_{\rm eff})$ used in Section~\ref{sec:security}.

\subsubsection{Weak Turbulence ($\sigma_R^2 < 1$)}
For weak turbulence (\(\sigma_R^2 < 1\)), the phase structure function \(D_\phi(r)\) represents the mean squared phase difference between two points separated by radial distance \(r = |\vec{\rho}|\) in the transverse plane:
\begin{equation}
D_\phi(r) = 1.09\, k^2 z \langle C_n^2 \rangle\, r^{5/3}.
\end{equation}
Here, the prefactor $1.09$ arises from a path-weighted averaging in spherical wave geometry, $z$ is the propagation distance, quantifying the intensity of the turbulence. This power-law form (\(r^{5/3}\)) reflects the inertial subrange of Kolmogorov turbulence. Equivalently, using the Fried parameter $r_0$,
\begin{align}
D_\phi(r)
&= 6.88\left(\frac{r}{r_0}\right)^{5/3}, \\
r_0^{-5/3}
&= 0.423\,k^2 \int_{0}^{z} C_n^2(\xi)\,{\rm d}\xi,
\end{align}
so that
\begin{equation}
D_\phi(r)
= 6.88\,r^{5/3}\,r_0^{-5/3}
= 2.91\,k^2\,r^{5/3}\!\int_{0}^{z} C_n^2(\xi)\,{\rm d}\xi.
\end{equation}
For a spherical wave, the standard geometric weighting leads to an extra factor $3/8$, turning $2.91$ into $2.91\times\frac{3}{8}=1.09$, i.e.
\begin{equation}
D_\phi(r)=1.09\,k^2\,r^{5/3}\,z\,\langle C_n^2\rangle,
\end{equation}
which matches the prefactor used above.
The rotation angle distribution is Gaussian, modeling small perturbations:
\begin{equation}
p(\theta\,|\,r) = \frac{1}{\sqrt{2\pi D_\phi(r)}} \exp\left(-\frac{\theta^2}{2 D_\phi(r)}\right).
\end{equation}
The depolarization factor is
\begin{equation}
\lambda(r) = \frac{1 - \exp\left(-\frac{D_\phi(r)}{2}\right)}{2},
\end{equation}

and decoherence factor
\begin{equation}
\bar{r}^2(r,\kappa) = 1 - \lambda(r) \frac{3\mu_\parallel(\kappa) - 1}{2},
\end{equation}
where $\mu_\parallel(\kappa) = \frac{1}{\kappa} \coth\kappa - \frac{1}{\kappa^2}$ is the axial second moment of the Fisher--Watson distribution (detailed in Appendix~\ref{app:phase_derivation}).

To obtain the final state, we average over the Gaussian beam profile and drift. The shifted intensity profile is
\begin{equation}
I(\vec{\rho} - \vec r_{\text{drift}}) = \frac{2}{\pi w_z^2} \exp\left(-\frac{2 |\vec{\rho} - \vec r_{\text{drift}}|^2}{w_z^2}\right).
\end{equation}
In polar coordinates \(\vec{\rho} = (r, \varphi)\), \(\vec r_{\text{drift}} = (r_{\text{drift}}, 0)\)):
\begin{equation}
|\vec{\rho} - \vec r_{\text{drift}}|^2 = r^2 + r_{\text{drift}}^2 - 2r r_{\text{drift}} \cos\varphi.
\end{equation}
Azimuthal integration leads to
\begin{equation}
\int_{0}^{2\pi}\exp\!\left(\frac{4 r r_{\text{drift}}}{w_z^2}\cos\varphi\right)\mathrm{d}\varphi
=2\pi\,I_0\!\left(\frac{4 r r_{\text{drift}}}{w_z^2}\right),
\label{eq:bessel_id}
\end{equation}
with $I_0$ the modified Bessel function of the first kind. 
Since the element of the area in the polar coordinates is $\mathrm{d}^2\rho = r\,\mathrm{d}r\,\mathrm{d}\varphi$, \emph{radial weight} (the probability density of radius $r$) is the azimuthal marginal of the normalized intensity:
\begin{align}
&W(r,r_{\text{drift}})
\equiv \int_{0}^{2\pi} I(\vec{\rho}-\vec r_{\text{drift}})\; r\,\mathrm{d}\varphi \nonumber\\
&= \frac{2}{\pi w_z^2}\,r\,
\exp\!\left[-\frac{2(r^2+r_{\text{drift}}^2)}{w_z^2}\right]
\int_{0}^{2\pi}\exp\!\left(\frac{4 r r_{\text{drift}}}{w_z^2}\cos\varphi\right)\mathrm{d}\varphi \nonumber\\
&= \frac{4r}{w_z^2}\,
\exp\!\left[-\frac{2(r^2+r_{\text{drift}}^2)}{w_z^2}\right]
I_0\!\left(\frac{4 r r_{\text{drift}}}{w_z^2}\right).
\label{eq:Wradial}
\end{align}
By construction$\int_{0}^{\infty} W(r,r_{\text{drift}})\,\mathrm{d}r=1$, so $W$ is a normalized radial pdf.
The aperture-limited integration (\(r \in [0, a]\)) gives
\begin{equation}
\lambda_a(r_{\text{drift}}) = \frac{\int_0^a W(r, r_{\text{drift}})\, \lambda(r)\, dr}
{\int_0^a W(r, r_{\text{drift}})\, dr},
\end{equation}
\begin{equation}
r_a^2(r_{\text{drift}}) = \frac{\int_0^a W(r, r_{\text{drift}})\, \bar{r}^2(r,\kappa)\, dr}
{\int_0^a W(r, r_{\text{drift}})\, dr}.
\end{equation}
\textcolor{black}{Similarly}, drift averaging produces effective parameters:
\begin{equation}
\lambda_a^{\text{eff}} = \int_0^\infty \frac{r_{\text{drift}}}{\sigma_{\text{drift}}^2}
\exp\left(-\frac{r_{\text{drift}}^2}{2 \sigma_{\text{drift}}^2}\right)
\lambda_a(r_{\text{drift}})\, dr_{\text{drift}},
\end{equation}
\begin{equation}
r_a^{2,\text{eff}} = \int_0^\infty \frac{r_{\text{drift}}}{\sigma_{\text{drift}}^2}
\exp\left(-\frac{r_{\text{drift}}^2}{2 \sigma_{\text{drift}}^2}\right)
r_a^2(r_{\text{drift}})\, dr_{\text{drift}}.
\end{equation}
The final detected state is
\begin{equation}
\hat{\rho}' = (1 - \lambda_a^{\text{eff}})
\begin{pmatrix}
\rho_{HH} & \rho_{HV} r_a^{2,\text{eff}} \\
\rho_{VH} r_a^{2,\text{eff}} & \rho_{VV}
\end{pmatrix}
+ \frac{\lambda_a^{\text{eff}}}{2} \mathbb{I},
\end{equation}
with total state
\begin{equation}
\hat{\rho}'_{\rm total}=\eta_{\rm total}\,\hat{\rho}'+(1-\eta_{\rm total})\,|\text{e}\rangle\!\langle\text{e}|
\end{equation}

The Gaussian law for $p(\theta|r)$ reflects a small-angle approximation consistent with $D_\phi(r)\ll 1$ over the aperture. In near-aperture saturation (large $r$ relative to $w_z$), the aperture-limited kernel $W(r,r_{\rm drift})$ ensures the proper weighting of partially decorrelated zones.
Detailed derivations are provided in Appendix~\ref{app:phase_derivation}.

\subsubsection{Medium Turbulence ($1 < \sigma_R^2 < 5$)}
For medium turbulence (\(1 < \sigma_R^2 < 5\)), phase fluctuations intensify, prompting a phenomenological phase structure function that combines Kolmogorov scaling with saturation:
\begin{align}
D_\phi(r) &= 2.22\, \sigma_R^2 \left( 1 - \exp\left( -\frac{r^2}{\text{r}_0^2} \right) \right) \left( 1 - \frac{\sigma_R^2}{5} \right) \notag \\
&\quad + 1.09\, k^2 z \langle C_n^2 \rangle\, r^{5/3} \cdot \frac{\sigma_R^2}{5},
\end{align}
The mixing parameter \(\alpha = 1 - \sigma_R^2/5\) interpolates the contributions. The rotation angle distribution mixes Gaussian and uniform:
\begin{equation}
p(\theta \mid r) = \alpha \frac{1}{\sqrt{2\pi D_\phi(r)}} \exp\left(-\frac{\theta^2}{2 D_\phi(r)}\right) + (1 - \alpha) \frac{1}{2\pi}.
\end{equation}
The uniform part fully depolarizes:
\begin{equation}
\int_0^{2\pi} \frac{1}{2\pi} U \hat{\rho} U^\dagger d\theta = \frac{\mathbb{I}}{2}.
\end{equation}
The depolarization and decoherence factors become:
\begin{equation}
\lambda(r) = 1 - \alpha \frac{1 + \exp\left(-\frac{D_\phi(r)}{2}\right)}{2}, \qquad
\bar{r}^2(r,\kappa) = \alpha \bar{r}'^2(r,\kappa),
\end{equation}
where \(\lambda'(r)\) and \(\bar{r}'^2(r,\kappa)\) follow the weak turbulence forms (detailed in Appendix~\ref{app:phase_derivation}). The final state follows the weak turbulence averaging, capturing the shift toward stronger effects via partial saturation. 
The mixture parameter $\alpha=1-\sigma_R^2/5$ is a phenomenological interpolant that enforces continuity between the weak-law and a Haar-averaged (fully depolarizing) limit. It should be regarded as a conservative surrogate rather than a unique physical law. A data-driven calibration of $\alpha(\sigma_R^2)$ and $\kappa(\sigma_R^2)$ can be obtained by fitting to wave-optics simulations or polarimetric field measurements; our simulations in Section~\ref{sec:simulations} instantiate only weak-turbulence profiles ($\sigma_R^2<1$), while the medium/strong forms are included here for completeness.

\subsubsection{Strong Turbulence ($\sigma_R^2 > 5$)}
For strong turbulence (\(\sigma_R^2 > 5\)), fluctuations overwhelm, yielding a uniform rotation angle distribution:
\begin{equation}
p(\theta \mid r) = \frac{1}{2\pi},
\end{equation}
leading to complete depolarization:
\begin{equation}
\mathcal{E}_{\text{phase},\,\vec{\rho}}(\hat{\rho}) = \frac{\mathbb{I}}{2}.
\end{equation}
The final state is
\begin{equation}
\hat{\rho}' = \frac{\mathbb{I}}{2}, \qquad 
\hat{\rho}'_{\rm total}=\eta_{\rm total}\,\hat{\rho}'+(1-\eta_{\rm total})\,|\text{e}\rangle\!\langle\text{e}|.
\end{equation}

The SU(2) Haar average, which leads to $\mathbb{I}/2$, is a worst-case upper bound on depolarization. Real links may retain residual coherence due to finite apertures, partial path correlation, and anisotropy ($\kappa>0$). Using the Haar limit in the security analysis is thus conservative. Detailed derivations are provided in Appendix~\ref{app:phase_derivation}.
We have developed a rigorous quantum channel framework to model atmospheric turbulence effects on polarization states in free-space MDI-QKD. By mapping phase perturbations to random SU(2) rotations, we quantify depolarization and decoherence through density matrices, deriving effective parameters \(\lambda_a^{\text{eff}}\) and \(r_a^{2,\text{eff}}\) that are experimentally measurable. The sequential channels--phase perturbation, beam spreading, drift, and aperture truncation--mirror physical propagation, ensuring consistency. Scintillation modulates only \(\eta_{\text{eff}}\), simplifying the analysis, while von Kármán, Hufnagel-Valley models, and corrected Rytov variance (Appendix~\ref{app:turbulence_models}) characterize regimes. Experimentally validated and compatible with security, this framework is a baseline for MDI-QKD optimization in turbulent channels, advancing reliable free-space quantum communication. For Bell-state measurement impacts, see Appendix~\ref{app:hom_visibility}. (Our numerical simulations instantiate only weak-turbulence profiles (clear/overcast/hazy) within $\sigma_R^2<1$, while the medium/strong forms are included here for completeness.)

\section{Security Analysis and Secret Key Rate}\label{sec:security}
Building on the depolarizing--dephasing description in Section~\ref{sec:channel}, we derive closed-form bounds on Eve's information and the resulting SKR for polarization-encoded MDI-QKD. The effects induced by turbulence are captured through the effective parameters \(\lambda_A = \lambda_a^{\text{eff},A}\), \(\lambda_B = \lambda_a^{\text{eff},B}\), \(\bar{r}^2 = r_a^{2,\text{eff}}\) (assuming symmetric channels for \(\bar{r}^2\)), and detection probabilities \(\eta_{\text{total}}^A\), \(\eta_{\text{total}}^B\). Integrating with the thermal loss and phase noise analysis in \cite{b38}, we adapt the framework to our turbulence model, derive key expressions for the total gain of the Z-basis \(Q_Z\), the effective key gain \(Q_Z^{1,1}\), the QBER of the Z-basis \(E_Z\), the QBER of the X-basis \(e_X^{1,1}\), and the SKR.

The total gain of the Z basis, the effective gain of the Z basis for key generation, the QBER of the Z basis, and the QBER of the X base are given by \cite{b38}:
\begin{align}
Q_Z &= \frac{1}{2} \eta_{\text{total}}^A \eta_{\text{total}}^B,\\
 Q_Z^{1,1} &= (2 - \lambda_A - \lambda_B + \lambda_A \lambda_B) \eta_{\text{total}}^A \eta_{\text{total}}^B,\\
 E_Z &= \frac{\lambda_A + \lambda_B - \lambda_A \lambda_B}{2},\\
 e_X^{1,1} &= \frac{1 - (1 - \lambda_A)(1 - \lambda_B) \bar{r}^4}{2}.  
\end{align}
For completeness, the key rate per emitted signal satisfies the Devetak-Winter bound:
\begin{equation}
R = I(A:B) - \chi(B:E) = Q_{Z}^{1,1} [1 - H(e_{X}^{1,1})] - Q_Z f H(E_Z)
\label{Annexeq:R_intermediate}
\end{equation}
Using the Devetak-Winter bound under reverse reconciliation \cite{b38}, the SKR is:
\begin{align}
R &= \eta_{\text{total}}^A \eta_{\text{total}}^B \biggl[ (2 - \lambda_A - \lambda_B + \lambda_A \lambda_B) \notag \\
&\quad \cdot \left( 1 - H\left( \frac{1 - (1 - \lambda_A)(1 - \lambda_B) \bar{r}^4}{2} \right) \right) \notag \\
&\quad - \frac{1}{2} f \, H\left( \frac{\lambda_A + \lambda_B - \lambda_A \lambda_B}{2} \right) \biggr],
\label{eq:skr}
\end{align}
where \(H(x) = -x \log_2 x - (1-x) \log_2 (1-x)\) is the binary entropy and \(f\) is the error correction inefficiency.
This analytical framework integrates turbulence-induced depolarization and decoherence into MDI-QKD security analysis, producing closed-form SKR expressions that facilitate numerical simulations in Section~\ref{sec:simulations} to evaluate performance under various distances and turbulence regimes.

\begin{figure*}[t!]
\centerline{\includegraphics[width=1\linewidth]{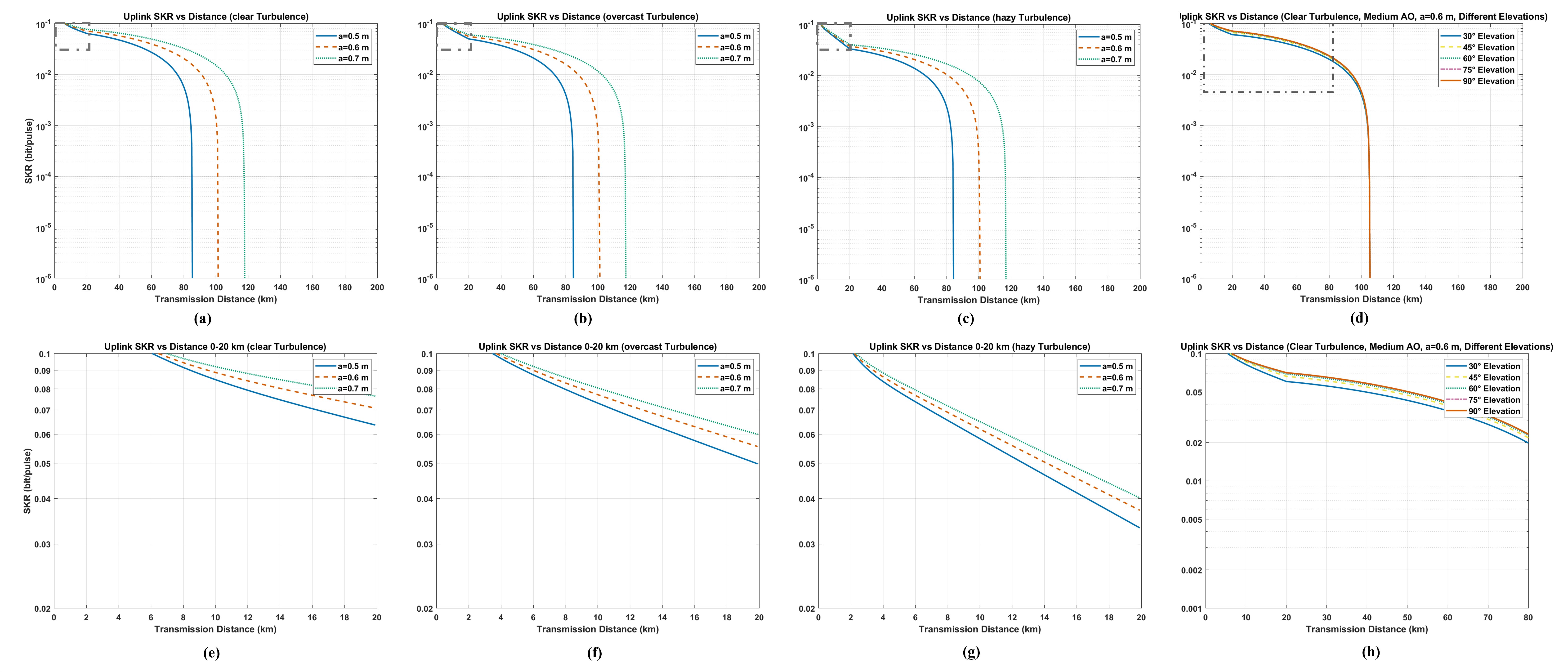}}
\caption{Uplink SKR versus transmission distance, combining aperture/turbulence sweeps and elevation dependence.
Panels (a)–(c): full range (0–200 km) under clear, overcast, and hazy turbulence, respectively, for receiver aperture radii $a\in\{0.5,0.6,0.7\}$\,m.
Panels (e)–(g): corresponding short-range zooms (0–20 km).
Panels (d) and (h): elevation dependence under clear turbulence with medium AO and $a=0.6$\,m, shown over 0–200 km and 0–80 km, respectively.
Legends indicate aperture radius in (a–c, e–g) and elevation angle in (d, h). Axes as labeled.}
\label{fig:skr_vs_distance_weather_elevation_uplink}
\end{figure*}

\section{Numerical Simulation and Results}
\label{sec:simulations}
\begin{table}[t]
\centering
\caption{Simulation parameters and weather--turbulence mapping.}
\label{tab:params}
\begin{tabular}{@{}l l@{}}
\toprule
\multicolumn{2}{@{}l}{\textbf{Optics \& Propagation}}\\
\midrule
Wavelength $\lambda$ & $850~\mathrm{nm}$ (Micius band) \\
Wavenumber $k$ & $2\pi/\lambda$ \\
Transmitter waist $w_0$ & $0.1~\mathrm{m}$ \\
Rayleigh range $z_R$ & $\pi w_0^2/\lambda$ \\
Atmospheric layer $H_{\rm atm}$ & $20~\mathrm{km}$ (turbulence truncation) \\
Elevation $\theta$ & $85^\circ$ (uplink unless stated) \\
Aperture radius set $a$ & $\{0.5,\,0.6,\,0.7\}~\mathrm{m}$ \\
\midrule
\multicolumn{2}{@{}l}{\textbf{Detector \& Electronics}}\\
\midrule
SNSPD quantum efficiency $\eta_{\rm det}$ & $0.75$ \\
Optical insertion loss & $2~\mathrm{dB}$ ($L_{\rm opt}=10^{-2/10}$) \\
Dark count rate $\xi_0$ & $1000~\mathrm{s}^{-1}$ (per SPD) \\
Gate width $\tau_G$ & $1~\mathrm{ns}$ \\
Dark-count per pulse $Y_0$ & $\xi_0 \tau_G$ \\
\midrule
\multicolumn{2}{@{}l}{\textbf{Numerics \& SKR}}\\
EC inefficiency $f$ & $1.1$ \\
QBER cutoff & $E_{\max}=15\%$ (set $R{=}0$ if $E_Z > E_{\max}$) \\
\bottomrule
\end{tabular}

\begin{tabular}{@{}l l@{}}
\toprule
\multicolumn{2}{@{}l}{\textbf{Weather-to-parameter mapping (uplink)}}\\
\midrule
Clear& $A=1.7{\times}10^{-14}~\mathrm{m}^{-2/3}$,\; $v=21~\mathrm{m/s}$,\; $\alpha_{\rm atm}=0.004~\mathrm{km}^{-1}$ \\
Overcast& $A=50{\times}10^{-14}$,\; $v=25~\mathrm{m/s}$,\; $\alpha_{\rm atm}=0.010~\mathrm{km}^{-1}$ \\
Haze& $A=100{\times}10^{-14}$,\; $v=30~\mathrm{m/s}$,\; $\alpha_{\rm atm}=0.020~\mathrm{km}^{-1}$ \\
\bottomrule
\end{tabular}

\begin{tabular*}{\columnwidth}{@{\extracolsep{\fill}}ll@{}}
\toprule
\multicolumn{2}{@{}l}{\textbf{Adaptive Optics (AO) baselines}}\\
\midrule
Mild   & $\rho_{\rm trk}{=}0.80,\; \kappa_w{=}0.95,\; \kappa_\phi{=}0.80$ \\
Medium & $\rho_{\rm trk}{=}0.50,\; \kappa_w{=}0.80,\; \kappa_\phi{=}0.60$ \\
Strong & $\rho_{\rm trk}{=}0.20,\; \kappa_w{=}0.60,\; \kappa_\phi{=}0.40$ \\
\bottomrule
\end{tabular*}

\begin{tabular*}{\columnwidth}{@{\extracolsep{\fill}}ll@{}}
\toprule
\multicolumn{2}{l}{\textbf{Decoy-state settings (three intensities, symmetric A/B)}}\\
\midrule
Signal intensity $\mu$ & $0.50$ \\
Decoy intensity $\nu$ & $0.10$ \\
Vacuum intensity & $0$ \\
Selection probabilities & $p_\mu=0.80,\; p_\nu=0.15,\; p_0=0.05$ \\
Basis bias & $p_Z=0.90,\; p_X=0.10$ \\
\bottomrule
\label{tab:decoy}
\end{tabular*}
\label{tab:decoy}
\end{table}

All simulations in this section consider an \emph{uplink} (ground-to-satellite), unless otherwise stated. With the parameter set in Table~\ref{tab:params}, we evaluate the composite channel of Section~\ref{sec:channel} on a grid over distance, aperture, weather class, elevation, and AO (precompensation) profiles. For each configuration we follow a fixed pipeline: (i) compute the Rytov variance $\sigma_R^2$ with uplink path-weighting (ground-layer dominated) and select the turbulence regime; (ii) evaluate the phase structure function to obtain local depolarization $\lambda(r)$ and decoherence $\bar r^2(r,\kappa)$; (iii) perform spatial averaging under Gaussian illumination with drift to obtain $\lambda_a^{\mathrm{eff}}$ and $r_a^{2,\mathrm{eff}}$; (iv) compute the end-to-end detection efficiency by combining aperture/beam-drift collection, atmospheric attenuation, optical/detector efficiencies, and scintillation (Gamma–Gamma); (v) assemble gains and QBER and evaluate the SKR using \eqref{eq:skr}, setting $R=0$ whenever the QBER cutoff in Table~\ref{tab:params} is exceeded. Unless explicitly varied in the figures, the baseline AO profile from Table~\ref{tab:params} is used.

Figure~\ref{fig:skr_vs_distance_weather_elevation_uplink} shows the uplink SKR versus distance for three weather classes and $a\!\in\!\{0.5,0.6,0.7\}$\,m. Within the first 0–20\,km, turbulence is the dominant impairment in uplink because the weighting concentrates near the transmitter; stronger weather leads to a steeper SKR decay, while larger apertures provide a consistent advantage through improved collection. Beyond 20\,km, after the turbulent layer is traversed, atmospheric attenuation $L_{\text{atm}}(\theta)$ (Section~\ref{sec:attenuation}) becomes the primary limiter and curves of a given weather class become nearly parallel on the log scale. Residual undulations stem from spatial averaging of broadened and drifted beams, which feed into $\lambda_a^{\mathrm{eff}}$ and $r_a^{2,\mathrm{eff}}$. The elevation panel (d) and (h) shows that higher elevation shortens the turbulent path and reduces the zenith-angle factor, giving a noticeable SKR improvement in the first tens of kilometers and a delayed cut-off at long range.

\begin{figure*}[t!]
\centerline{\includegraphics[width=1\linewidth]{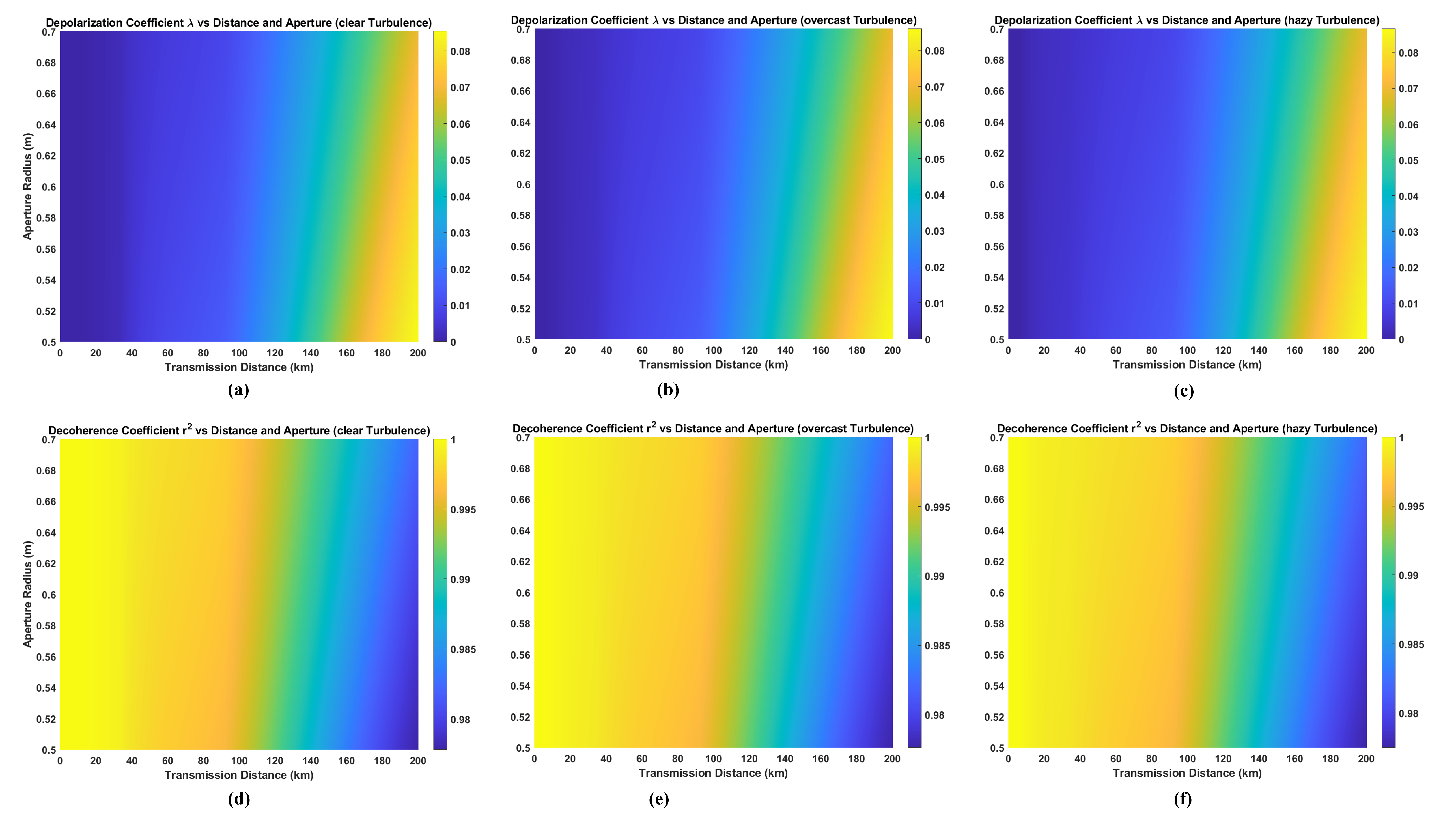}}
\caption{Uplink depolarization coefficient $\lambda$ (top row) and decoherence coefficient $r^2$ (bottom row) versus transmission distance and receiver aperture radius, with adaptive optics (AO). Range: 0–200 km.
(a),(d) Clear; (b),(e) Overcast; (c),(f) Hazy.
Axes: horizontal—distance (km); vertical—aperture radius $a$ (m).
Color bars indicate the value of $\lambda$ and $r^2$, respectively.}
    \label{fig:lambda_r2_AO_full_uplink}
\end{figure*}

\begin{figure}
\centering
\includegraphics[width=\columnwidth]{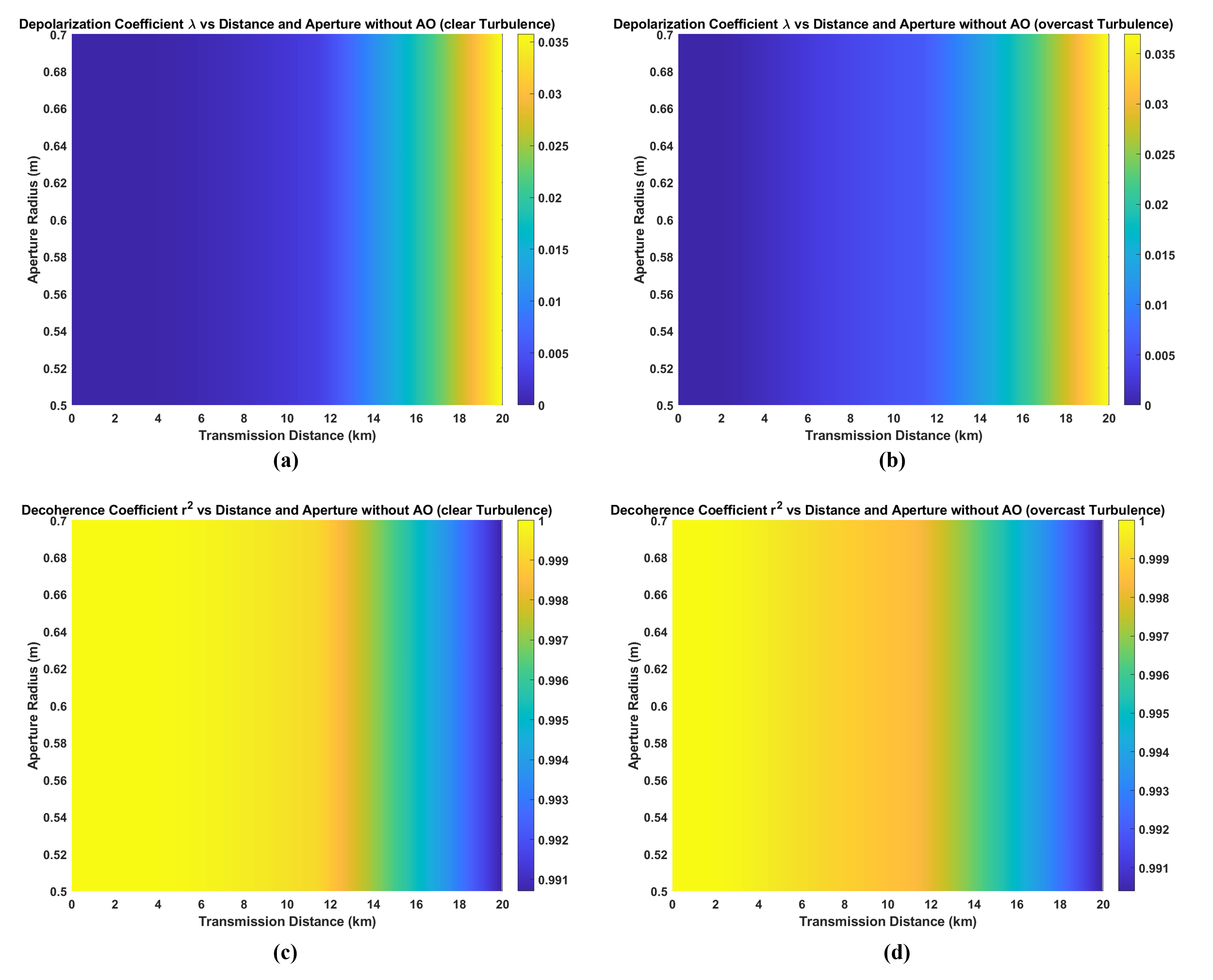}  
\caption{Uplink depolarization coefficient $\lambda$ (top) and decoherence coefficient $r^2$ (bottom) versus distance and aperture, \emph{without} AO, short range (0–20 km).
Panels: (a)–(b) $\lambda$ for clear/overcast; (c)–(d) $r^2$ for clear/overcast.
Axes: horizontal—distance (km); vertical—aperture radius $a$ (m).}  \label{fig:lambda_r2_noAO_short_uplink}

\end{figure}

The effective depolarization and decoherence parameters are visualized in Fig.~\ref{fig:lambda_r2_AO_full_uplink} (with AO, 0–200\,km) and Fig.~\ref{fig:lambda_r2_noAO_short_uplink} (without AO, 0–20\,km). With AO under clear/overcast conditions, $\lambda_a^{\mathrm{eff}}$ remains small and $r_a^{2,\mathrm{eff}}$ close to unity over a broad range, indicating a modest mixing of polarization in the instantiated weak-turbulence profiles. Without AO, both $\lambda_a^{\mathrm{eff}}$ and $r_a^{2,\mathrm{eff}}$ degrade faster with distance due to uncompensated phase distortions and beam wander; in those cases SKR is curtailed mainly by QBER growth.

\begin{figure}
\centering
\includegraphics[width=\columnwidth]{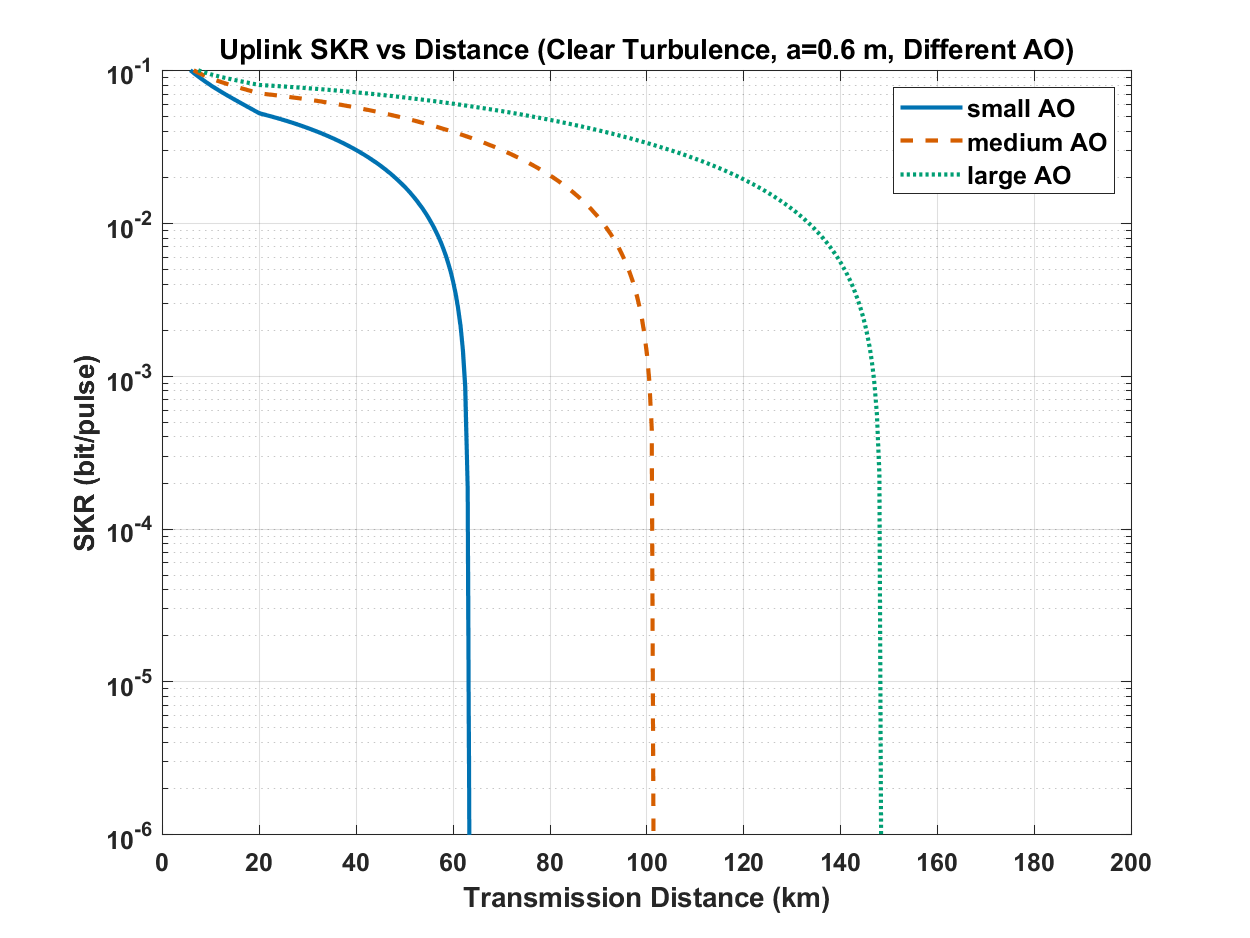}  
\caption{Effect of AO strength on uplink SKR (clear turbulence, $a=0.6$ m). 
Curves compare ``small'', ``medium'', and ``large'' AO profiles.
Axes: horizontal—distance (km); vertical—SKR (bit/pulse, log scale).}  \label{fig:skr_AO_strength_uplink}
\end{figure}

Figure~\ref{fig:skr_AO_strength_uplink} quantifies the benefit of stronger AO for a fixed aperture ($a=0.6$, m) in clear weather: Moving from 'small' to 'large' AO shifts the cutoff to much larger distances and produces orders of magnitude gains in the midrange. 

\begin{figure}
\centering
\includegraphics[width=0.8\columnwidth]{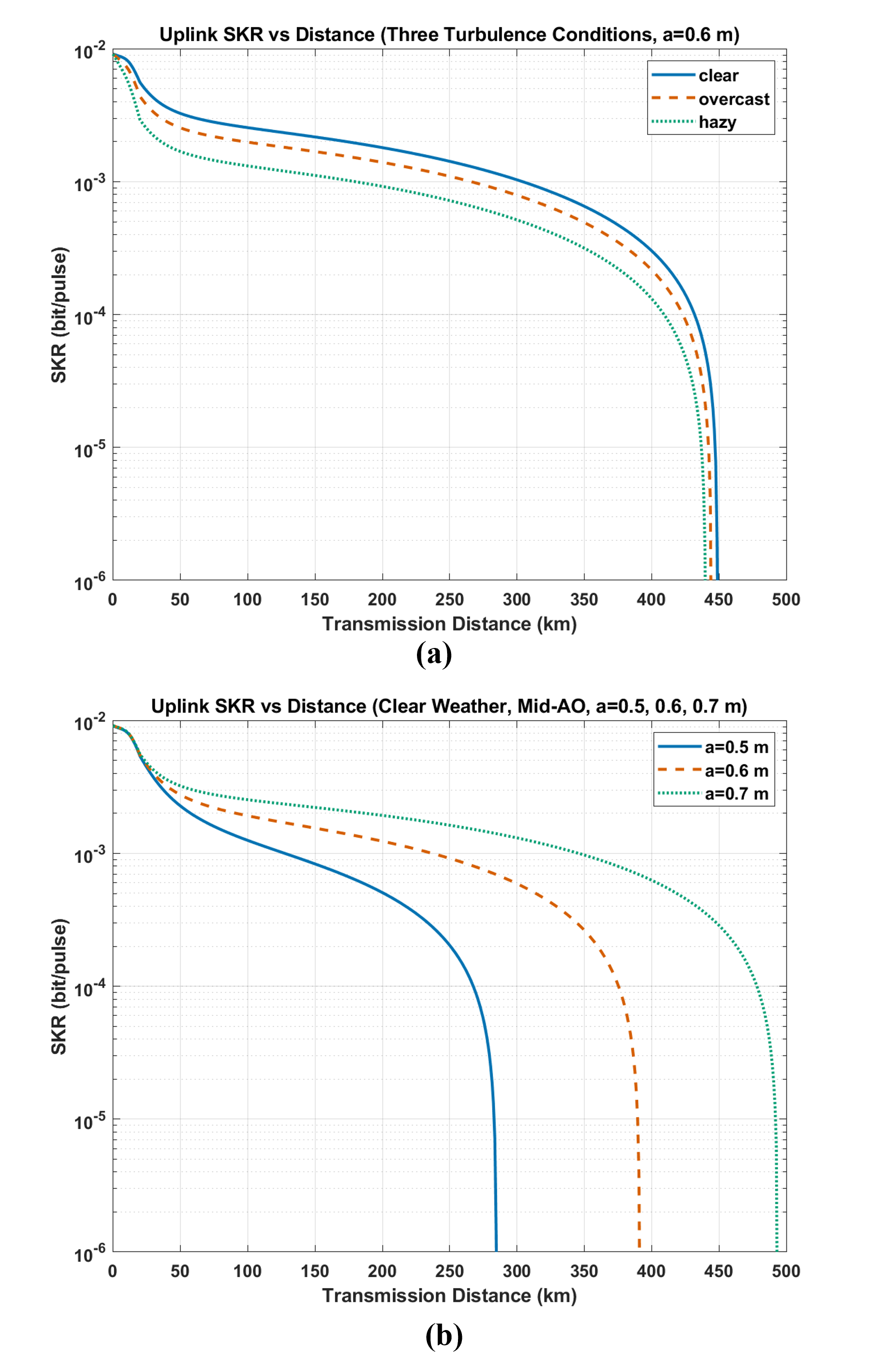}  
\caption{Uplink decoy-state MDI-QKD SKR vs distance (a) clear/overcast/haze at 0.6m aperture; (b) $a\!\in\!\{0.5,0.6,0.7\}$ m }  
\label{fig:decoy_skr}
\end{figure}
Using the three-intensity Decoy state MDI-QKD setting of Table~\ref{tab:decoy} and the estimator in
Appendix~\ref{app:decoy}, Fig.~\ref{fig:decoy_skr} summarizes uplink SKR under the
composite channel. Panel (a) compares clear/overcast/hazey at $a{=}0.6$\,m (medium AO):
the curves follow the expected weather ordering with a short-range “knee’’ where
turbulence dominates, then an attenuation-limited decay. Panel (b) varies the aperture
in clear weather: a larger $a$ monotonically raises SKR and extends the cutoff range.
Both level and slope are consistent with the MDI-QKD analyses of the decoy state reported for
FSO links and with the QKD trends from space to ground~\cite{b6,b39,b40}, supporting the
validity of the prediction of the composite channel.

Overall, the simulations validate that the composite channel of Section~\ref{sec:channel} captures SKR trends across weather, aperture, elevation, and AO profiles: AO extends the viable SKR range under our parameter set (e.g., to 100 km in clear and 50 km in overcast for $a{=}0.7$\,m), while in the absence of AO the polarization channel rapidly collapses ($\lambda_a^{\mathrm{eff}}\!\to\!1$, $r_a^{2,\mathrm{eff}}\!\to\!0$), pushing QBER above the cutoff and nullifying SKR. 

Limitations of  Turbulence Regimes:  (i) The HV parameters $(A,v)$ and $\alpha_{\text{atm}}$ are site- and season-dependent; our weather mapping is a representative but not universal choice. 
(ii) The AO model is parametric (spread/drift scaling) and does not include closed-loop bandwidth or temporal lag; measured AO transfer functions can be incorporated in future work. 
(iii) The axis-preference parameter $\kappa$ (von Mises–Fisher/Watson) is phenomenological; empirical calibration versus slant path and wind shear would refine $r_a^{2,\mathrm{eff}}$. 
(iv) Pointing jitter, scintillation, and partial spatial-mode mismatch at the BSM are treated independently; a joint spatio-polarization model could tighten SKR bounds. 
(v) Strong-turbulence cases are rare under standard HV profiles; thresholds separating weak/medium/strong may be updated with site-specific campaigns.

\section{Conclusion}
\label{sec:conclusion}

We have introduced a compact, physically grounded composite channel model that distills the complex effects of atmospheric turbulence into an effective depolarizing--dephasing operation on polarization qubits. By sequentially incorporating phase perturbations, Gaussian beam spreading, beam drift, aperture truncation, and scintillation into a unified framework, our model mirrors the physical propagation dynamics while yielding closed-form expressions for key parameters such as the depolarization factor $\lambda_a^{\mathrm{eff}}$, decoherence factor $r_a^{2,\mathrm{eff}}$, and detection probability $\eta_{\mathrm{eff}}$. This enables one-shot security proofs for free-space MDI-QKD and provides analytic predictions of the SKR across weak, medium, and strong turbulence regimes, seamlessly integrating with existing analyses of thermal-loss and phase-noise channels\cite{b38}.

The closed-form nature of our model offers key advantages over traditional methods. Unlike computationally intensive Monte Carlo wave–optics simulations, which scale with propagation distance and aperture size, our phenomenological SU(2) rotation mapping—with Fisher–Watson–distributed axes and regime-specific angle distributions - enables lightweight real-time computations for dynamic link adaptation. For example, it supports rapid parameter sweeps (e.g., zenith angle or aperture radius) to assist ground-to-satellite link design under temporal turbulence variations. \textbf{We focus on the weak-scintillation regime} $(\sigma_R^2 \le 1)$; Unless explicitly stated, all derivations and quantitative claims are calibrated in this regime.

Compared to more precise methods such as multiphase-screen propagation, which resolve detailed wavefront distortions but demand high computational resources, our model trades granular accuracy for efficiency while retaining essential physics: capturing depolarization and decoherence with measurable parameters that can be calibrated against polarimetric measurements \emph{in the weak-scintillation regime}, and remaining consistent with known limiting cases (no turbulence and aperture-only). We do not include a head-to-head benchmark against wave–optics Monte Carlo; quantifying model bias relative to phase-screen simulations is left for future work. Limitations include the assumption of statistical independence of perturbations and simplified temporal correlations, which may underestimate effects in rapidly evolving turbulence or multipath scenarios. Additionally, reliance on phenomenological distributions may overlook higher-order anisotropies under extreme conditions.

Future work will address these points by validating the model against full-wave numerical simulations and field experiments on moving satellite platforms, incorporating real-time AO feedback. Extensions could include time-bin or higher-dimensional encodings for enhanced resilience, integration with diversity techniques, refinement of phase distributions via empirical data, and inclusion of temporal dynamics to capture turbulence evolution. Ultimately, this framework aims to enable scalable, secure quantum links over turbulent free-space channels while keeping computational cost low.

\appendices

\section{Turbulence Models and Rytov Variance}
\label{app:turbulence_models}
The statistical properties of atmospheric turbulence are critical to accurately modeling its effects on photon propagation, as discussed in Sections~\ref{sec:channel} and~\ref{sec:turbulence_regimes}. Here, we derive the foundational models used in the main text: the von Kármán spectrum for the spatial power spectral density of refractive index fluctuations and the Hufnagel-Valley (HV) model for altitude-dependent turbulence strength. These enable precise calculations of the phase structure function, beam drift variance, and Rytov variance, with corrections for Gaussian beam divergence to support the regime-specific parameters \(\lambda_a^{\text{eff}}\) and \(r_a^{2,\text{eff}}\).

\subsection{von Kármán Spectrum}
The von Kármán spectrum extends the Kolmogorov turbulence model by incorporating finite outer and inner scales, making it suitable for modeling both phase perturbations and beam drift across all turbulence regimes. The power spectral density of refractive index fluctuations is:
\begin{equation}
\Phi_n(\kappa) = \frac{0.033 C_n^2}{(\kappa^2 + \kappa_0^2)^{11/6}} \exp\left(-\frac{\kappa^2}{\kappa_m^2}\right), \quad \kappa_0 = \frac{2\pi}{L_0}, \quad \kappa_m = \frac{5.92}{l_0}, \label{eq:von_karman}
\end{equation}
where \(\kappa\) is the spatial wave number (in \(\text{m}^{-1}\)), \(C_n^2\) is the refractive index structure constant (in \(\text{m}^{-2/3}\)), \(L_0 = 25 \, \text{m}\) is the outer scale of turbulence and \(l_0 = 0.001 \, \text{m}\) is the inner scale. The constant 0.033 derives from Kolmogorov theory, adjusted for the von Kármán form, and the exponent \(-11/6\) captures the power-law scaling in the inertial subrange. The exponential term suppresses high-frequency contributions, ensuring realism for small-scale eddies.
This spectrum underpins the phase structure function \(D_\phi(r)\) in the weak and medium regimes (see Appendix~\ref{app:phase_derivation}) and the beam drift variance \(\sigma_{\text{drift}}^2\) in Section~\ref{sec:beam_drift}. For divergent Gaussian beams that approximate spherical waves (\(z \gg z_R\)), we integrate the tilt spectrum to obtain the following.
\begin{equation}
\sigma_{\text{drift}}^2 = 0.73 \langle C_n^2 \rangle z^3 w_z^{-1/3} \left[1 - 0.97 \left(\frac{L_0}{w_z}\right)^{1/3}\right] \sec^2\beta,
\end{equation}
where the prefactor 0.73 reflects spherical-wave geometry, and the correction term ensures non-negative variance by accounting for large-eddy averaging when \(w_z > L_0\). The factor \(\sec^2\beta\) adjusts for the elongation of the path at the zenith angle \(\beta\), directly supporting the drift channel \(\E{drift}\) in the main text.

\subsection{Hufnagel-Valley Model}
The Hufnagel-Valley (HV) model provides the altitude-dependent refractive index structure constant \(C_n^2(h)\), essential for path-averaged calculations in satellite links:
\begin{align}
C_n^2(h) &= 0.00594 \left( \frac{v}{27} \right)^2 (10^{-5} h)^{10} \exp\left( -\frac{h}{1000} \right) \notag \\
&\quad + 2.7 \times 10^{-16} \exp\left( -\frac{h}{1500} \right) + A \exp\left( -\frac{h}{100} \right), \label{eq:hv_model_app}
\end{align}
where \(h\) is altitude (in meters), \(v = 21 \, \text{m/s}\) is wind speed, and \(A = 1.7 \times 10^{-14} \, \text{m}^{-2/3}\) is ground-level turbulence strength. The terms model high-altitude wind shear, stratospheric contributions, and ground-layer turbulence, respectively. The path averaged \(\langle C_n^2 \rangle\) is:
\begin{equation}
\langle C_n^2 \rangle = \frac{1}{z} \int_0^z C_n^2(h(\xi)) d\xi, \label{eq:c_n2_avg}
\end{equation}
with truncation at \(z = 20 \, \text{km}\) for satellite links, such as \(C_n^2(h > 20 \, \text{km}) \approx 0\). This truncation aligns with the atmospheric thickness in Section~\ref{sec:attenuation}, preventing the overestimation of turbulence in the upper layers and supporting realistic values \(\sigma_R^2\) in simulations (Section~\ref{sec:simulations}).
\subsection{Rytov Variance}
The Rytov variance $\sigma_R^2$ quantifies the turbulence strength and is commonly used to classify regimes as weak $(<1)$, medium $(1\!-\!5)$, or strong $(>5)$. For a slant path of length $z$ at elevation $\theta$ (zenith angle $\beta=\tfrac{\pi}{2}-\theta$), we parameterize the optical path by $\xi\in[0,z]$ with local altitude $h(\xi)=\xi\sin\theta$. A beam-size–corrected spherical-wave expression is
\begin{align}
&\sigma_R^2(z,\beta)
= 2.25\,k^{7/6}\sec^{11/6}\!\beta \notag\\
&\qquad\times \int_0^{\min(z,H_{\rm atm})}
 C_n^2\!\big(h(\xi)\big)\, W_{\rm link}(\xi,z)\,
 F_{\rm beam}(\xi)\, d\xi,
\label{eq:rytov_uplink_dl}
\end{align}
\begin{align}
W_{\rm link}(\xi,z)
&=
\begin{cases}
(\xi/z)^{5/6}\big(1-\xi/z\big)^{5/6}, & \text{downlink},\\[2pt]
\big(1-\xi/z\big)^{5/6},               & \text{uplink},
\end{cases}
\\[2pt]
F_{\rm beam}(\xi)
&=\Big[1+\big(\tfrac{w_0}{w_z(\xi)}\big)^2\Big]^{-7/6}.
\end{align}
where $k=2\pi/\lambda$ and $w_z(\xi)=w_0\sqrt{1+(\xi/z_R)^2}$. For constant $C_n^2$ this reduces to the familiar approximation $\sigma_R^2 \approx 0.4\, C_n^2 k^{7/6} z^{11/6}$. The truncation at $H_{\rm atm}\!\approx\!20$ km reflects that $C_n^2(h)$ is negligible above the turbulent layer.

The phase-structure integral used for coherence metrics and the Fried parameter employs a related kernel:
\begin{align}
D_{\rm int}(z,\beta)
&= \int_0^{\min(z,H_{\rm atm})}
 C_n^2\!\big(h(\xi)\big)\, W_D(\xi,z)\,
 F_D(\xi)\, d\xi,
\label{eq:Dint_uplink_dl}
\end{align}
\begin{align}
W_D(\xi,z)
&=
\begin{cases}
\big(1-\xi/z\big)^{5/3}, & \text{downlink},\\[2pt]
(\xi/z)^{5/3},           & \text{uplink},
\end{cases}
\\[2pt]
F_D(\xi)
&=\Big[1+\big(\tfrac{w_0}{w_z(\xi)}\big)^2\Big]^{-1}.
\end{align}
In all simulations, we adopt the \emph{uplink} kernels in \eqref{eq:rytov_uplink_dl}–\eqref{eq:Dint_uplink_dl}; these emphasize ground-layer contributions, consistent with uplink geometry.

\section{Phase Channel Derivation}
\label{app:phase_derivation}
The phase-perturbation channel $\mathcal{E}_{\text{phase},\,\vec{\rho}}$ maps
turbulence-induced wavefront phase fluctuations to local polarization rotations at transverse position $\vec{\rho}$, as introduced in Section~\ref{sec:phase_channel}.
For concreteness, we detail the derivation in the weak-turbulence regime; the same averaging steps (axis distribution, angle statistics, and spatial averaging) apply to other regimes and lead to the depolarizing--dephasing form used to define the effective parameters $\lambda_a^{\mathrm{eff}}$ and $r_a^{2,\mathrm{eff}}$ after spatial averaging in Section~\ref{sec:turbulence_regimes}.

\subsection{Weak Turbulence with Anisotropic Rotation Axes}
For $\sigma_R^2<1$, small phase fluctuations lead to a Gaussian rotation angle $\theta$ conditioned on $r=|\vec{\rho}|$:
\begin{equation}
p(\theta \mid r)
= \frac{1}{\sqrt{2\pi D_\phi(r)}}\,
  \exp\!\Big(-\tfrac{\theta^2}{2 D_\phi(r)}\Big),
\label{eq:weak_p_theta}
\end{equation}
with phase structure function
\begin{equation}
D_\phi(r)
= 1.09\,k^2 z \,\langle C_n^2\rangle\, r^{5/3},
\label{eq:weak_dphi}
\end{equation}
where $1.09$ accounts for spherical-wave averaging.
The rotation axis follows the Fisher--Watson distribution
\begin{equation}
w_\kappa(\phi)
= \frac{\kappa}{2\sinh\kappa}\,e^{\kappa\cos\phi}\,\sin\phi,\quad
\kappa\ge 0,
\label{eq:fisher_watson}
\end{equation}
with concentration $\kappa$ favoring the propagation direction.
The SU(2) rotation operator is
\begin{equation}
U(\theta,\phi,\alpha)
= \cos\!\Big(\tfrac{\theta}{2}\Big)\,\mathbb{I}
  - i \sin\!\Big(\tfrac{\theta}{2}\Big)\,(\vec{n}\cdot\vec{\sigma}),
\label{eq:rotation_op}
\end{equation}
with$\vec{n}=(\cos\phi,\;\sin\phi\cos\alpha,\;\sin\phi\sin\alpha)$. Axis moments read
\begin{equation}
\mu_\parallel(\kappa)=\langle n_z^2\rangle
= \frac{1}{\kappa}\coth\kappa - \frac{1}{\kappa^2},\qquad
\mu_\perp(\kappa)=\frac{1-\mu_\parallel}{2}.
\label{eq:mu_moments}
\end{equation}

Averaging over the rotation axis $(\phi,\alpha)$ gives
\begin{equation}
\big\langle U \hat{\rho} U^\dagger \big\rangle_{\phi,\alpha}
= \Big(\cos^2\!\tfrac{\theta}{2}+\mu_\parallel\sin^2\!\tfrac{\theta}{2}\Big)\hat{\rho}
+ 2\mu_\perp\sin^2\!\tfrac{\theta}{2}\,
\mathrm{diag}(\rho_{VV},\,\rho_{HH}).
\end{equation}
Subsequent averaging over $\theta$ leads to
\begin{equation}
\lambda(r)=\frac{1-e^{-D_\phi(r)/2}}{2},\qquad
\bar{r}^2(r,\kappa)=1-\lambda(r)\,\frac{3\mu_\parallel(\kappa)-1}{2},
\label{eq:r_lambda}
\end{equation}
and the local channel
\begin{equation}
\mathcal{E}_{\text{phase},\,\vec{\rho}}(\hat{\rho})
=(1-\lambda(r))
\begin{bmatrix}
\rho_{HH} & \rho_{HV}\,\bar{r}^2(r,\kappa)\\
\rho_{VH}\,\bar{r}^2(r,\kappa) & \rho_{VV}
\end{bmatrix}
+\frac{\lambda(r)}{2}\,\mathbb{I}.
\label{eq:phase_channel}
\end{equation}

%

\subsection{CPTP and Conditional Post-Selection}
Let $K_{\theta,\vartheta,\varphi}
= \sqrt{p(\theta\mid r)\,w_\kappa(\vartheta)\,\tfrac{1}{2\pi}}\;U(\theta,\vec{n}).
$
Since $\int p(\theta\mid r)\,d\theta=1$, $\int w_\kappa(\phi)\,d\Omega=1$, and $\alpha$ is uniform on $[0,2\pi)$,
\begin{equation}
\int d\theta\,d\Omega\,
K_{\theta,\phi,\alpha}^\dagger K_{\theta,\phi,\alpha}
=\mathbb{I},
\end{equation}
so $\mathcal{E}_{\text{phase},\,\vec{\rho}}$ is CPTP. Aperture truncation and detection form a classical post-selection: the resulting (renormalized) polarization map is CP but generally not TP, consistent with the treatment in the main text.

\section{Spatial Indistinguishability and HOM Visibility}
\label{app:hom_visibility}
In MDI-QKD, HOM interference for Bell-state measurements requires spatial mode overlap, potentially reduced by turbulence-induced drift (Section~\ref{sec:beam_drift}). Gaussian profiles from Alice/Bob are:
\begin{equation}
I_{A,B}(\vec{\rho}) = \frac{2}{\pi w_z^2} \exp\left(-\frac{2|\vec{\rho} - \vec r_{\text{drift},A,B}|^2}{w_z^2}\right).
\end{equation}
Overlap is:

\begin{equation}
\begin{aligned}
    \langle M_{\text{spatial}} \rangle &= \left| \int d^2\vec{\rho} \sqrt{I_A(\vec{\rho}) I_B(\vec{\rho})} \right|^2\\& = \exp\left( -\frac{|(\vec r_{\text{drift}})_A - (\vec r_{\text{drift}})_B|^2}{w_z^2} \right).
\end{aligned}
\end{equation}

Averaging over Rayleigh-distributed drift (\(\sigma_{\text{drift}}^2\)) gives:
\begin{equation}
\langle M_{\text{spatial}} \rangle = \frac{1}{1 + 2 \sigma_{\text{drift}}^2 / w_z^2}.
\end{equation}

For typical links (\(\sigma_R^2 \lesssim 1\), \(w_0 = 0.3~\text{m}\)), \(\sigma_{\text{drift}} \lesssim 0.1 w_z\), so \(\langle M_{\text{spatial}} \rangle \gtrsim 0.98\), justifying the approximation to 1 unless there are extreme conditions. For example, with \(\sigma_R^2 = 1\), \(z = 20~\text{km}\), \(\langle C_n^2 \rangle = 10^{-14}~\text{m}^{-2/3}\), \(\sigma_{\text{drift}}^2 \approx 0.005 w_z^2\), yielding \(\langle M_{\text{spatial}} \rangle \approx 0.99\). However, if \(\sigma_{\text{drift}} > 0.5 w_z\) (e.g., strong turbulence without AO), \(\langle M_{\text{spatial}} \rangle < 0.8\), invalidating the approximation and requiring spatial mode corrections. This supports the channel model's focus on polarization effects, assuming that HOM visibility remains high in weak-medium regimes. If spatial indistinguishability is not ideal, the HOM visibility reduction $\,\langle M_{\rm spatial}\rangle\le1\,$ effectively scales the polarization coherence entering the X basis. A conservative replacement consistent with Sec.~\ref{sec:security} is
\begin{equation}
e_X^{1,1}=\frac{1-(1-\lambda_A)(1-\lambda_B)\big(\langle M_{\rm spatial}\rangle\,\bar r^2\big)^2}{2}.
\end{equation}
In our simulations we adopt $\langle M_{\rm spatial}\rangle\simeq1$ (Appendix.~\ref{app:hom_visibility}); the above provides a correction rule whenever severe drift or misalignment is present.

\section{Scintillation Effect on Polarization}
\label{app:scintillation}
Scintillation modulates intensity via Gamma-Gamma distribution (Section~\ref{sec:scintillation}):
\begin{equation}
p(I_0) = \frac{2 (\alpha \beta)^{\frac{\alpha + \beta}{2}}}{\Gamma(\alpha) \Gamma(\beta)} I_0^{\frac{\alpha + \beta}{2} - 1} K_{\alpha - \beta}\left(2 \sqrt{\alpha \beta I_0}\right), \label{eq:gamma_gamma}
\end{equation}
with \(\alpha,\beta\) as in the main text. The final state with scintillation and drift is
\begin{equation}
\begin{aligned}
\hat{\rho}_{\text{final}}
&= \frac{1}{\eta_{\text{eff}}}
   \int_0^\infty p(I_0)\, I_0\, \mathrm{d}I_0
   \int_0^\infty p_{\text{drift}}(r)\, \eta(r)\,
\\
&\qquad \times
   \Bigg[
   (1-\lambda_a(r))
   \begin{bmatrix}
     \rho_{HH} & \rho_{HV}\, r_a^2(r)\\
     \rho_{VH}\, r_a^2(r) & \rho_{VV}
   \end{bmatrix}
   + \frac{\lambda_a(r)}{2}\,\mathbb{I}
   \Bigg]\,
   \mathrm{d}r .
\end{aligned}
\label{eq:rho_final}
\end{equation}
where \(\eta(r),\lambda_a(r),r_a^2(r)\) are defined in the main text, and
where \(\eta(r),\lambda_a(r),r_a^2(r)\) are defined in the main text, and
\begin{equation}
\eta_{\text{eff}}
= \int_0^\infty\! p(I_0)\,I_0\,\mathrm d I_0
  \int_0^\infty\! p_{\text{drift}}(r)\,\eta(r)\,\mathrm d r.
\label{eq:eta_eff_app}
\end{equation}
Because \(I_0\) enters only as a scalar multiplier, it cancels in normalization:
scintillation modulates \(\eta_{\text{eff}}\), but not the polarization block.
This holds under (i) polarization-insensitive coupling/collection, (ii) linear detector response within the gate, and (iii) no polarization-dependent loss after turbulence; otherwise the separation in \eqref{eq:rho_final}–\eqref{eq:eta_eff_app} becomes approximate.

\section{Decoy-State MDI-QKD Security Analysis and SKR}
\label{app:decoy}

The numerical study in Section~\ref{sec:simulations} employs three-intensity decoy states; here, we provide a self-contained derivation of the secret key rate (SKR) for MDI-QKD under the composite free-space turbulence channel developed in Section~\ref{sec:channel}. The goal is to expose where the turbulence physics enters the decoy analysis while keeping the standard estimators intact.

The polarization dynamics of the detected ensemble are summarized by the spatially averaged depolarization and decoherence coefficients: $\lambda_a^{eff}(z), $$ \text{r}_a^{2,\text{eff}}(z)$,as obtained in Section~\ref{sec:turbulence_regimes}. Scintillation is modeled by a Gamma--Gamma intensity factor \(I_0\) of unit mean with shape parameters \(\alpha,\beta\) (Appendix~\ref{app:scintillation}). Writing the per-pulse transmissivity as \(\eta_\text{total}=\eta_{\text{eff}} I_0\), with \(\eta_{\text{base}}\) collecting aperture/beam-drift collection, atmospheric attenuation, and optics/detector efficiencies, the first two moments needed for MDI are
\begin{equation}
 \qquad
\langle\eta^2\rangle=\eta_{\mathrm{total}}^{(2)}=\eta_{\text{eff}}^{2}
\!\left(1+\frac{1}{\alpha}\right)\!\left(1+\frac{1}{\beta}\right).
\label{app:eq:moments}
\end{equation}
When both arms share the same scintillation statistics (symmetric uplink or common path segment), we use \(\langle\eta_A\eta_B\rangle=\eta_{\mathrm{total}}^{(2)}\); if the branches are independent, \(\langle\eta_A\eta_B\rangle=\eta_{\mathrm{total}}^2\).

Depolarization reduces the two-photon Hong–Ou–Mandel success for the Bell state measurement (BSM). The polarization-dependent BSM coefficient is
\begin{equation}
F\!\left(\lambda_a^{\mathrm{eff}}\right)
=\frac{2-2\lambda_a^{\mathrm{eff}}+(\lambda_a^{\mathrm{eff}})^2}{4},
\label{app:eq:F}
\end{equation}
which recovers \(1/2\) in the non-depolarization limit. The corresponding single-photon error on the basis of X, including decoherence, is
\begin{equation}
e_{11}^{X}
=\frac{1-\bigl(1-\lambda_a^{\mathrm{eff}}\bigr)^2\bigl(r_a^{2,\mathrm{eff}}\bigr)^2}{2}.
\label{app:eq:e11x}
\end{equation}
Any residual system mismatch (e.g., temporal or spatial mode visibility) can be absorbed as a multiplicative visibility factor if needed, but is not required for the results reported here.

Let \(Y_0\) denote the probability of dark count per gate. The elementary leads to vacuum and single-photon input follow directly from \eqref{app:eq:moments} and \eqref{app:eq:F}:
\begin{equation}
\begin{aligned}
Y_{00}&=Y_0^2\\
Y_{10}&=Y_{01}=Y_0\,\langle\eta\rangle\\
Y_{11}&=F\!\left(\lambda_a^{\mathrm{eff}}\right)\,\langle\eta_A\eta_B\rangle .
\label{app:eq:leads:to}
\end{aligned}
\end{equation}
Vacuum or single-arm events err with probability \(1/2\); the single-photon pair errs with probability \(e_{11}^{X}\) from \eqref{app:eq:e11x}. With Poissonian sources of intensities \(x,y\in\{\mu,\nu,0\}\) and weights \(P_n(x)=e^{-x}x^n/n!\), the observed gains and error gains in basis \(\text{B}\in\text{Z},\text{X}\}\) are
\begin{equation}
\begin{aligned}
Q_{x,y}^{\mathsf{B}}
&= \sum_{n,m\ge 0} P_n(x) P_m(y)\, Y_{nm}^{\mathsf{B}},\\
E_{x,y}^{\mathsf{B}} Q_{x,y}^{\mathsf{B}}
&= \sum_{n,m\ge 0} P_n(x) P_m(y)\, e_{nm}^{\mathsf{B}} Y_{nm}^{\mathsf{B}}.
\end{aligned}
\label{app:eq:Q-Exy}
\end{equation}
For three-intensity decoys with \(\mu>\nu>0\), it is convenient to define \(S_{xy}=e^{x+y}Q_{x,y}\) and \(T_{xy}=e^{x+y}E_{x,y}Q_{x,y}\). The standard tight estimators then give the single-photon-pair yield lower bound and X-basis error upper bound as
\begin{align}
Y_{11}^{L}
&=\frac{\mu}{\mu\nu(\mu-\nu)}\!\left(S_{\nu\nu}-S_{\nu0}-S_{0\nu}+S_{00}\right)\nonumber\\
&\quad-\frac{\nu}{\mu\nu(\mu-\nu)}\!\left(S_{\mu\mu}-S_{\mu0}-S_{0\mu}+S_{00}\right),
\label{app:eq:Y11L}\\
e_{11}^{X,U}
&=\frac{T_{\nu\nu}-T_{\nu0}-T_{0\nu}+T_{00}}{\nu^2\,Y_{11}^{L}},
\qquad 0\le e_{11}^{X,U}\le\tfrac12.
\label{app:eq:e11U}
\end{align}
In the numerical implementation, one clips \(Y_{11}^{L}\) at zero to avoid spurious negativity due to finite precision.

Finally, the asymptotic, symmetric-basis SKR follows the Devetak–Winter form with decoy estimates inserted. Writing \(P_1(\mu)=\mu e^{-\mu}\), with the Z-basis selection probability \(p_Z\), the signal intensity probability \(p_\mu\) and error correction inefficiency \(f\),
\begin{equation}
R \;\ge\; p_Z^2\,p_\mu^2\Big[
P_1(\mu)^2\,Y_{11}^{\mathsf Z,L}\bigl(1-H(e_{11}^{\mathsf X,U})\bigr)
- f\,Q_{\mu,\mu}^{\mathsf Z}\,H\!\bigl(E_{\mu,\mu}^{\mathsf Z}\bigr)
\Big],
\label{app:eq:R-decoy}
\end{equation}
where \(Y_{11}^{\mathsf Z,L}\) and \(e_{11}^{\mathsf X,U}\) are given by \eqref{app:eq:Y11L}–\eqref{app:eq:e11U}, and \(H(\cdot)\) is the binary entropy as defined in Section~\ref{sec:security}. 

\end{document}